\documentclass[a4paper, total={6in, 8in}, left=2cm, right=2cm, top=2cm, bottom=2cm]{article}
\usepackage{multirow}
\usepackage{blindtext}
\usepackage{wrapfig}
\usepackage{graphicx} 
\usepackage[utf8]{inputenc}
\usepackage[english]{babel}

\usepackage[backend=biber, style=ieee, sorting=none]{biblatex}
\usepackage{comment}
\usepackage{array}
\graphicspath{{figures/}}
\usepackage{caption}
\usepackage{subcaption}
\usepackage{algorithm}
\usepackage{algpseudocode}
\usepackage{amsmath}
\usepackage{amssymb}
\usepackage{tocloft}    
\usepackage{ragged2e}   
\usepackage{hyperref}   
\usepackage{glossaries}
\usepackage{booktabs}
\usepackage{arxiv}

\newenvironment{conditions}
  {\par\vspace{\abovedisplayskip}\noindent\begin{tabular}{>{$}l<{$} @{${}={}$} l}}
  {\end{tabular}\par\vspace{\belowdisplayskip}}

\newcommand{\listequationsname}{\Large{List of Equations}}
\newlistof{myequations}{equ}{\listequationsname}
\newcommand{\myequations}[1]{
   \addcontentsline{equ}{myequations}{\protect\numberline{\theequation}#1}
}
\title{Exploring the impact of reflexivity theory and cognitive social structures on the dynamics of doctor-patient social system}

\author{ \href{https://orcid.org/0000-0002-7976-9140}{\includegraphics[scale=0.06]{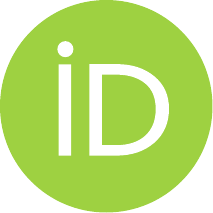}\hspace{1mm}Al Saqib Majumder}\\
	Department of Engineering and Informatics\\
	University of Sussex\\
	Brighton, BN19BJ \\
	\texttt{am2597@sussex.ac.uk} \\
}

\renewcommand{\shorttitle}{Exploring the impact of reflexivity theory and cognitive social structures}

\hypersetup{
pdftitle={Exploring the impact of reflexivity theory and cognitive social structures on the dynamics of doctor-patient social system},
pdfsubject={cs.SI, cs.NE},
pdfauthor={Al Saqib ~Majumder},
pdfkeywords={Social systems, Social networks, Cognitive social structures, Reflexivity theory},
}

\begin{document}

\maketitle

\begin{abstract}
	Conventional economic and socio-behavioural models assume perfect symmetric access to information and rational behaviour among interacting agents in a social system. However, real-world events and observations appear to contradict such assumptions, leading to the possibility of other, more complex interaction rules existing between such agents. We investigate this possibility by creating two different models for a doctor-patient system. One retains the established assumptions, while the other incorporates principles of reflexivity theory and cognitive social structures. In addition, we utilize a microbial genetic algorithm to optimize the behaviour of the physician and patient agents in both models. The differences in results for the two models suggest that social systems may not always exhibit the behaviour or even accomplish the purpose for which they were designed and that modelling the social and cognitive influences in a social system may capture various ways a social agent balances complementary and competing information signals in making choices.  
\end{abstract}


\section{Introduction}

Conventional economic and behavioural studies have long assumed that an agent has complete access to information and adheres to rational behaviour to model interactions between social agents \protect\cite{KorobkinUlen2000, Soros2013}. However, evidence challenging these assumptions has accumulated, pointing to other, more plausible theories \protect\cite{BromileyPapenhausen2003, Crotty2017}. This paper presents a novel model of social system design and analysis. Our model incorporates ideas from the theory of reflexivity \cite{Soros2013} and cognitive social structures \cite{Krackhardt1987a} offering a socio-cognitive approach to understanding the behaviours of individual social agents and emergent social phenomena. 

\begin{figure}[h!]
    \centering
    \includegraphics[width=1\linewidth]{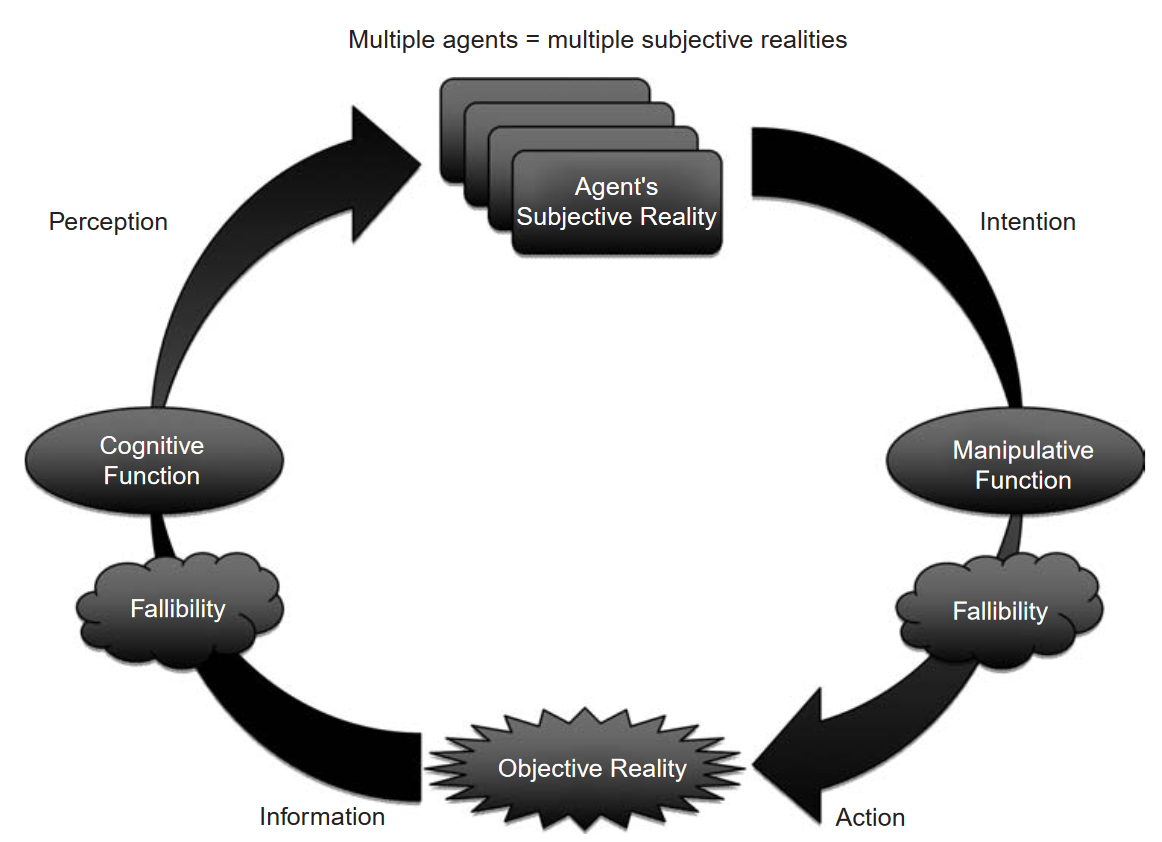}
    \caption{Outline of Reflexivity Theory \protect\cite{Soros2013}}
    \label{fig:Outline of Reflixivity Theory}
\end{figure}

Reflexivity theory, considered formally proposed for economic analysis by George Soros \cite{Soros2013},\cite{Umpleby2018}, is similar to the concepts found in the studies and literature of second-order cybernetics \cite{Scott2004}. The theory, in a nutshell, suggests that social agents and the environment, which constitute a single system, are involved in a feedback loop with not just negative feedback, as usually considered to be the case in classical economic analysis of markets and social phenomena, but also positive feedback to and from the agents and the environment driving the overall system towards a specific, attractor state \cite{Davis2020}. The critical insight from the theory that we have utilized in our model is that feedback loops between the agents and their environments change both the agents and the environments; once the agents act, the environments that they are in change as well, which influences the following actions the agents take and so it continues. Furthermore, the social agents can only access their own subjective realities of these changes, which are different from an objective reality (See Figure \ref{fig:Outline of Reflixivity Theory}). These subjective realities can often drive many social phenomena, including business cycles and stock market bubbles \cite{Beinhocker2013},\cite{Soros2013}, despite their possible deviations from the objective reality.

Cognitive social structures, an extension of the theory of social structures, was instrumental in understanding that social agents are situated in a social network with relationship dynamics and that the position of social agents in a network influences the perceptions, decisions and actions of each agent in that network \cite{Brands2013},\cite{Krackhardt1987a}. Once again, similar to what we observe in the theory of reflexivity, the social agents do not have direct access to objective reality according to this framework. Instead, social agents perceive their interactions with others through a subjective lens, forming the basis for their actions and decisions. However, the critical difference between the two theories is that while the theory of reflexivity explores the dynamics of the interactions of social agents in a system, cognitive social structures uncover the matrix of relationships that influence the subjective perceptions of each agent \cite{Frank2015}. 

Our primary motivation behind pursuing this research is the intuition that socially embedded, reflexive agents have different sets of behaviours that lead to the emergence of social phenomena which are closer to what we observe in our social systems compared to what we observe from classical economic and sociological models where rationality and perfect information assumptions are made. It is essential to realize and recognize that perceptions of individual agents are real in their consequences, even if there is no direct, one-to-one relationship between observed behaviours and objective reality. To flesh out our intuition, we compare and contrast observations from the simulation of two types of models, which we call the "classical" and "cognitive social system", respectively. In particular, we explore the behaviours of doctors and patients in a primary healthcare network. We have chosen this social system partly because of the abundance of literature regarding the choices and behaviours of patients and doctors \cite{Djulbegovic.etal2014},\cite{Harris2003},\cite{Kozikowski.etal2022} and also because of its utilitarian nature; better comprehension of this system would lead to more robust and precise healthcare services that will help in saving lives and improving standards of living \cite{Cabrera.etal2011},\cite{Comis.etal2021}.

The classical model employs perfect information and rational behaviour assumptions, while the cognitive social system integrates the theory of reflexivity and cognitive social structures. We want to find out whether the best doctors, as determined by certain traits in the models, such as their credentials and abilities to conduct research, among others, get the most patients. In classical models, the best doctors would ideally have the best reputation because patients would know the best doctors and choose them accordingly to receive treatments. However, in the cognitive social system, the best doctors may not have the most favourable reputations because of social ties, which modulate the perception of their abilities and, thus, alter the judgements of patients choosing them. We implement a microbial genetic algorithm to optimize the agents in both models \cite{Harvey2011}. We use this variation of the genetic algorithm because both of our models are relatively small in scale, and using a complete, conventional genetic algorithm may detract from the focus of our analysis. While a microbial genetic algorithm is more skeletal than a conventional one, it is more than adequate for our purposes in this paper. Ultimately, we are convinced that cognitive social systems can be generalized to understand other social systems, such as those found in areas such as education and defense, among others, and even design new ones. The potential of cognitive social system modelling to capture the dynamics of agents' interactions embedded in social systems more accurately than conventional models is immense, as long as the modeller can rely on sound literature and verify the models with well-founded data.
\vspace{3mm}

\section{Methodology}

\begin{figure}[h!]
    \centering
    \includegraphics[width=1\linewidth]{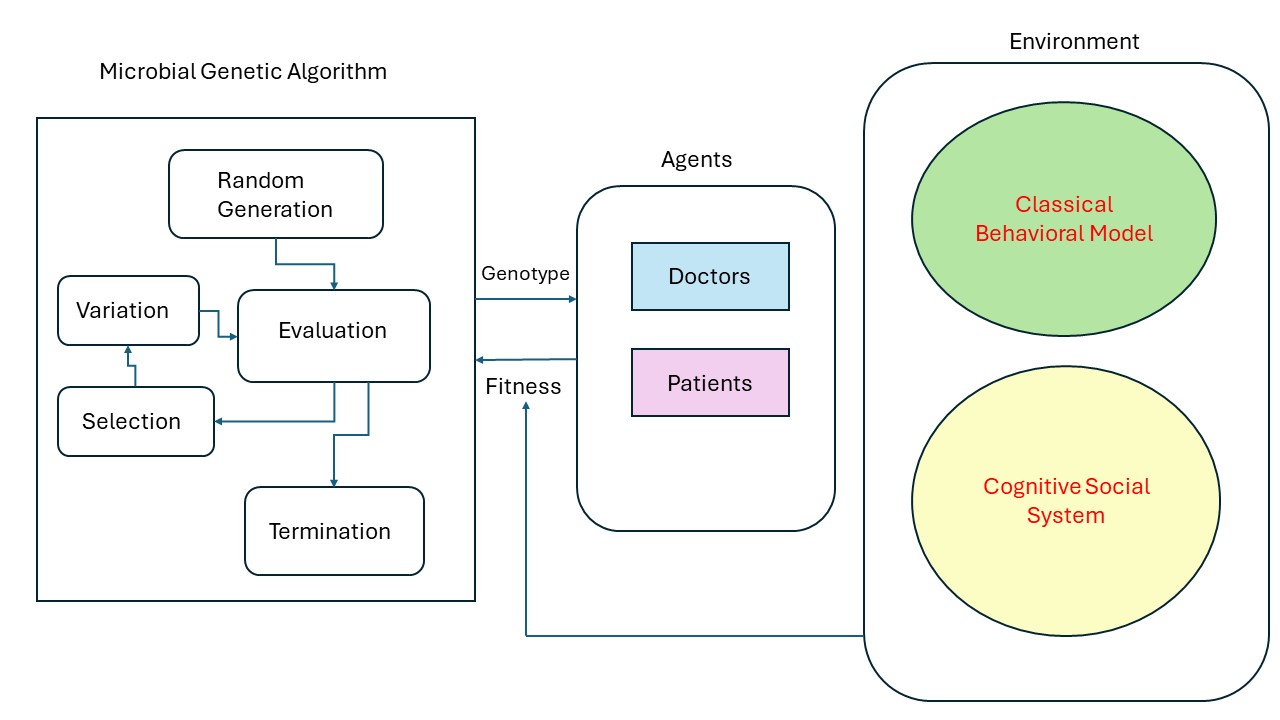}
    \caption{Overview of the Research Experiment Workflow.}
    \label{fig:Research_Experiment_workflow}
\end{figure}

\begin{figure}[h!]
    \centering
    \includegraphics[width=1\linewidth]{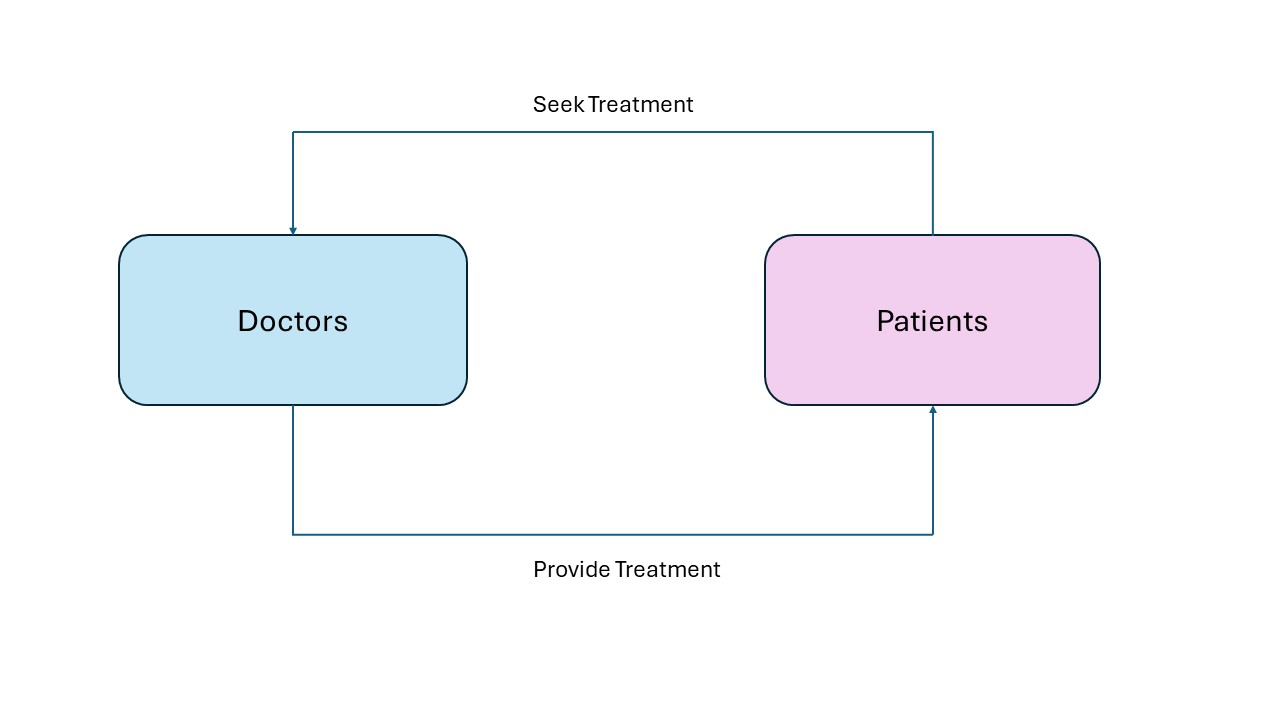}
    \caption{Agent Interaction Rule}
    \label{fig:interaction_rule}
\end{figure}

As discussed briefly in the introduction, the main focus of this paper involves modelling a doctor-patient healthcare system using two different approaches. Modelling the system for both frameworks involves a collection of patients requiring treatments from multiple doctors for infections. Thus, the healthcare system can be considered a social system of primary care networks, where patients can directly approach doctors for relevant treatments \cite{Comis.etal2021}. We utilise a microbial genetic algorithm to co-evolve patients' and doctors' attributes and characteristics in our models; the evolutionary process models learn along the generations of the agents \cite{Farago.etal2022}. Thus, the interaction of doctors and patients in our models can be seen as a form of biological cooperation, where both agents strive to optimise their complementary fitness functions. Learning across generations can be thought of as learning after each interaction between the agents, with the interaction being a form of repeated games \cite{Farago.etal2022}. Thus, implementing the genetic algorithm in this system can be interpreted through a more grounded socioeconomic lens. Figure \ref{fig:Research_Experiment_workflow} illustrates the overall experimental workflow for our research paper, and Figure \ref{fig:interaction_rule} depicts the general interaction rules between the agents we are interested in modelling. 

Our decision to focus on a primary healthcare network as our system of analysis was not arbitrary. We chose this system with careful consideration to ensure that the design of the components of our models would remain the main focus of our investigation. We also wanted to work with a system that was sufficiently complex without becoming too unpredictable. While we could have chosen other social systems and even different areas of the healthcare system such as emergency healthcare network \cite{Cabrera.etal2011} among others \cite{Tracy.etal2018}, we decided to model the primary healthcare network. This choice was driven by the fact that it mainly involved modelling the interactions of doctors and patients. The restricted range of components and interaction possibilities naturally lent itself to a simple agent-based modelling approach. We will discuss the implemented interactions and components in detail in the following paragraphs. 

\subsection{Classical Model}

\begin{figure}[h!]
    \centering
    \includegraphics[width=1\linewidth]{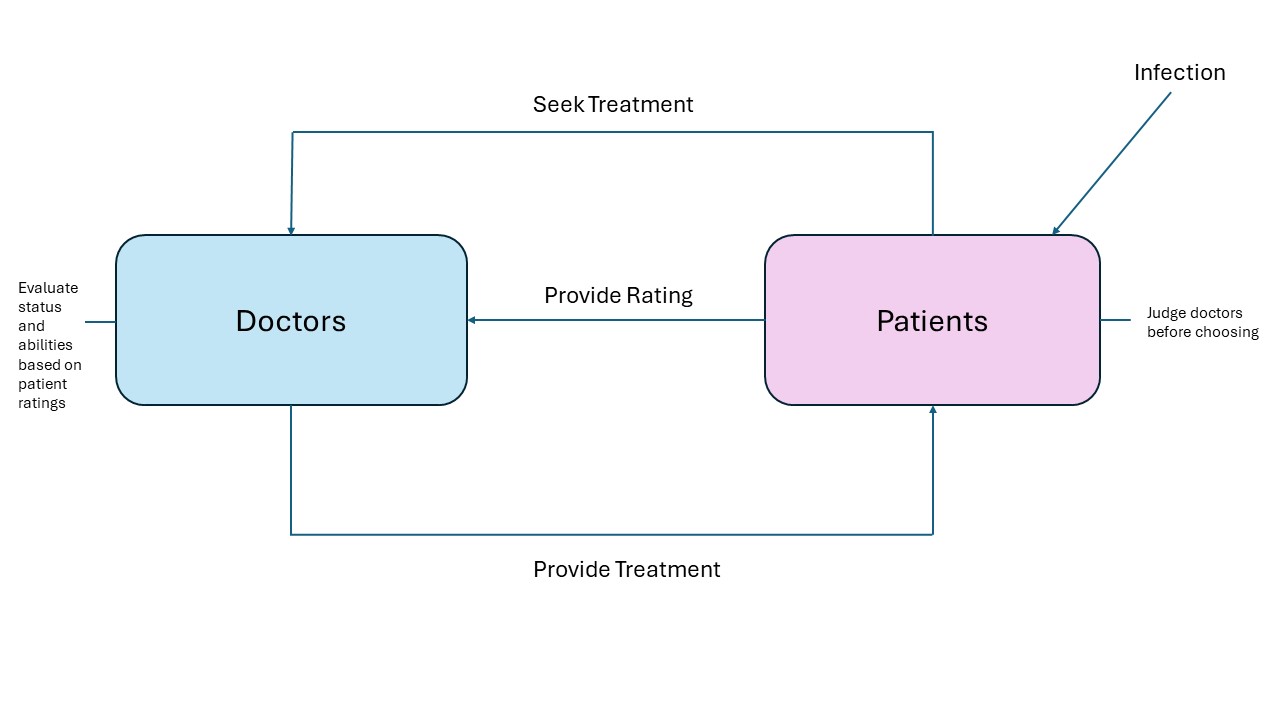}
    \caption{Classical Model Interaction Rule}
    \label{fig:classical_interaction_rule}
\end{figure}

In the classical model, as mentioned in the previous section, we assume that the social agents have perfect information and can optimise their objectives based on the feedback they receive from each other. In the model, the doctors want to maximise the ratings they receive from patients, and patients want to maximise their health levels. Patients have a health level between $0$ and $1$ in our model. Patients are infected randomly, with a fixed infection severity of $0.2$, which is the amount by which the health level of the patients decreases. Once treated by doctors, patients can be infected again after they recover. Patients search for a doctor when their health level is below $0.6$, and they continue to search for a doctor until their health level is $0.8$ or above. The patients receive feedback on their choices of doctors by noting how much of their health has been recovered from the treatments. We assume that patients can make choices regarding their doctors and that they can access the services without any restrictions. The health level of the patients can be thought of as an aggregate score of how well the vital organs are functioning through medical measurements such as blood pressure, x-ray scans and others \cite{Stratford.etal1996}. The doctors receive feedback for their treatment through ratings by patients, which depends on how well the doctors' treatments can help patients recover their health level. The ratings can be considered a signal of reputation and, therefore, desirability for the doctor propagated through word of mouth, online discussions, and even direct ratings on relevant medical websites \cite{Chen.Lee2024}. Figure \ref{fig:classical_interaction_rule} summarises the main interaction logic for the agents in the classical model. Except for the patient ratings, values for all the other properties for both agents are normalized to be between 0 and 1. 

\subsubsection{Doctor Agent}

The doctor agent in the model has multiple properties, which are called attributes and actions, also known as methods, for interacting with the patient agent. We will discuss the main attributes and methods essential for understanding the functionality and behaviour of the doctor agent in the model since there are other attributes and methods that exist in our program to make our code easier to access and read. The doctor treats a patient and improves the patient's health based on her abilities and credentials and receives a rating in return. Based on the rating, the doctor can decide how much change is necessary for her skills, if any at all. The main properties or attributes that the doctor agent has are as follows: 

\begin{enumerate}
    \item Experience (\texttt{experience}): It is mainly a counter that keeps track of how many patients a doctor has treated in the simulation. It is used as one of the factors in determining whether a doctor can improve her credential level in the model \cite{Patel.Sharma2024}.
 
    \item Personal Resource (\texttt{personal\_resource}): Personal resource determines how much a doctor can develop her skills, such as research ability and empathy, which are essential for improving her treatment effectiveness. It can be thought of as a cognitive resource that can be utilized for increasing either of the two attributes \cite{Vecchio1990}. It is fixed for every doctor to be $0.2$. 

    \item Technological Resource Constraint (\texttt{technological\_resource\_constraint}): It is a random value between $0.2$ and $0.5$, initialized differently for each doctor and represents the technological limitations a doctor faces, such as lack of diagnostic machines and proper tools to improve treatment for patients \cite{Eklund2008}. It directly affects the level of treatment effectiveness a doctor can possess at any given moment in the simulation. 

    \item Research Ability (\texttt{research\_ability}): It is an ability of a doctor that represents the doctor's research endeavours and, therefore, the ability to improve technical knowledge and capabilities \cite{Davies2007}. The doctor can improve this ability based on patient feedback; a higher rating means that the doctor does not need to improve her knowledge since it is adequate to treat the patients, while a lower rating means that the doctor needs to improve her knowledge capacity. 

    \item Empathy (\texttt{empathy}): Similar to research ability, a doctor can improve his empathetic ability based on patient feedback. This attribute was added because it is considered one of the most important abilities of doctors in enhancing treatment outcomes through psychological effects \cite{Hojat.etal2002}. 

    \item Credential (\texttt{credential}): It is divided into three categories, ranging from high, medium to low. A doctor can improve credential by improving her research ability and increasing her experience values. It is one of the most important factors that contribute to patients choosing the doctor \cite{Harris2003}, \cite{Kozikowski.etal2022}. 
\end{enumerate}

\begin{table}[h!]
\centering
\begin{tabular}{|l|p{8cm}|}
\hline
\textbf{Name in Code} & \textbf{Description} \\ \hline
\texttt{doctor\_id} & Unique identifier for each doctor. \\ \hline
\texttt{experience} & Tracks the experience level of the doctor, incremented after treating each patient. \\ \hline
\texttt{research\_ability} & A random value initially set between 0.2 and 0.6, representing the doctor's ability to conduct and apply research. \\ \hline
\texttt{empathy} & A random value initially set between 0.2 and 0.7, representing how empathetic the doctor is towards patients. \\ \hline
\texttt{personal\_resource\_constraint} & Represents a fixed constraint on the doctor's ability to develop their skills, initially set to 0.8. \\ \hline
\texttt{personal\_resource} & Calculated as 1 minus the personal resource constraint (0.8), representing the resources available for the doctor's development. \\ \hline
\texttt{technological\_resource\_constraint} & A random value between 0.2 and 0.5 that affects the effectiveness of treatments provided by the doctor. \\ \hline
\texttt{credential} & Represents the qualification level of the doctor, initially set randomly to either 'low', 'medium', or 'high'. \\ \hline
\end{tabular}
\caption{Main Properties of the Doctor Agent in the Classical Model}
\label{tab:DoctorClassicalProperties}
\end{table}

The doctor also has a property called \texttt{effectiveness}, which is calculated through one of the methods, \texttt{calculate\_treatment\_effectiveness}. Although technically not an attribute, it is responsible for determining by how much a patient's health level improves after the patient is treated by the doctor. It is dependent on the values of doctor's empathy, credential and technological resource constraint. Table \ref{tab:DoctorClassicalProperties} provides a summary of the properties of the doctors in this model. The main actions or methods that a doctor has are:

\begin{enumerate}
    \item Treating a patient (\texttt{treat\_patient}): This method determines how a doctor treats a patient. It utilizes treatment effectiveness calculation and experience and research update calculations to complete the treatment procedure that is implemented. Pseudocode \ref{code:treat_patient} encapsulates the logic of the method. 
    
    \item Experience and Research Update Calculations (\texttt{update\_experience\_and\_abilties}):  The method determines how a doctor updates experiences and research abilities in the model, after treating the patient. The action incorporates upgrading credential method and adds a value of 1 for every patient treated by an individual doctor agent. Pseudocode \ref{code:update_experience} summarizes the method.

    \item Upgrading credential (\texttt{upgrade\_credential}): It follows a logic that requires research ability and experience of the doctor to increase to a certain level for the credential to improve. This method of updating credential follows a general idea of how credentials are improved by doctors in the physical world, with experience and research abilities both playing a part in increasing the desirability and respectability of the doctors \cite{Cassidy.etal2019}, \cite{Patel.Sharma2024}. The values for research ability are chosen to reflect that higher research ability is meant to represent the capacity for higher credential. Equation \ref{equation:cred_upgrade} and Pseudocode \ref{code:upgrade_credential} are provided to capture how this method is implemented precisely. 

    \begin{equation}
    \text{credential} = 
    \begin{cases} 
    \text{'medium'} & \text{if } r \geq 0.5 \text{ and } exp \geq 50 \text{ and credential = 'low'}\\
    \text{'high'} & \text{if } r \geq 0.8 \text{ and } exp \geq 80 \text{ and credential = 'medium'}\\
    \text{credential} & \text{otherwise}
    \end{cases}  
    \label{equation:cred_upgrade}
    \end{equation}
    where:
    \begin{conditions}
     r     &  research ability \\
     exp     &  experience \\   
    \end{conditions}
    \myequations{Credential Upgrade Logic}

    \item Treatment effectiveness calculation (\texttt{calculate\_treatment\_effectivenes}): This method calculates the treatment effectiveness that determine the increase in health level and therefore the rating received by the doctor from the patient. Both credential and empathy play a vital role in calculating the treatment effectiveness of the doctor, since both technical knowledge and psychological aspects of the doctor influence the patient's recovery \cite{Kozikowski.etal2022}. Equation \ref{equation:treat_effect_calc} and Pseudocode \ref{code:calculate_treat_effectiveness} depict how the method has been provided in the model. 

    The treatment effectiveness (\( E \)) based on the doctor's credentials is calculated as follows:
    \begin{equation}
    E = min (0.7, (C + E) \times (1 - TRC))
    \label{equation:treat_effect_calc}
    \end{equation}
    where:
    \begin{conditions}
     C     &  credential with low: $0.1$, med: $0.2$, high: $0.3$ \\
     E     &  empathy \\   
     TRC & technological\_resource\_constraint
    \end{conditions}
    \myequations{Treatment Effectiveness Equation}
\end{enumerate}

Table \ref{tab:MethodsDoctorClassical} provides an overview of the methods in the doctor agent class. Updating research ability and empathy of the doctor agent are handled by the genetic algorithm implemented in the simulation. A doctor agent also has an instantiation of the rating manager class, which is used to manage rating calculations in the model. 

\begin{algorithm}[h!]
\caption{Treat Patient}
\begin{algorithmic}[1]
\Function{treat\_patient}{}
    \If{\text{self.is\_busy}}
        \State \Return $0$
    \EndIf
    \State \Call{start\_treatment}{}
    \State $\text{effectiveness} \gets \Call{calculate\_treatment\_effectiveness}{}$
    \State \Call{update\_experience\_and\_abilities}{}
    \State \Call{end\_treatment}{}
    \State \Return $\text{effectiveness}$
\EndFunction
\end{algorithmic}
\label{code:treat_patient}
\end{algorithm}

\begin{algorithm} [h!]
\caption{Update Experience and Abilities}
\begin{algorithmic}[1]
\Function{update\_experience\_and\_abilities}{}
    \State $\text{self.experience} \gets \text{self.experience} + 1$
    \State \Call{upgrade\_credential}{}
\EndFunction
\end{algorithmic}
\label{code:update_experience}
\end{algorithm}

\begin{algorithm}[h!]
\caption{Upgrade Credential}
\begin{algorithmic}[1] 
    \Function{upgrade\_credential}{}
        \If{$\text{self.credential} = \text{'low'}$ \textbf{and} $\text{self.research\_ability} \geq 0.5$ \textbf{and} $\text{self.experience} \geq 50$}
            \State $\text{self.credential} \gets \text{'medium'}$
        \ElsIf{$\text{self.credential} = \text{'medium'}$ \textbf{and} $\text{self.research\_ability} \geq 0.8$ \textbf{and} $\text{self.experience} \geq 80$}
            \State $\text{self.credential} \gets \text{'high'}$
        \EndIf
    \EndFunction
\end{algorithmic}
\label{code:upgrade_credential}
\end{algorithm}

\begin{algorithm}[h!]
\caption{Calculate Treatment Effectiveness}
\begin{algorithmic}[1]
\Function{calculate\_treatment\_effectiveness}{}
    \State $\text{credential\_factor} \gets \Call{GetCredentialFactor}{\text{self.credential}}$
    \State $\text{effectiveness} \gets (\text{credential\_factor} + \text{self.empathy}) \times (1 - \text{self.technological\_resource\_constraint})$
    \State \Return $\min(\text{effectiveness}, 0.7)$
\EndFunction
\end{algorithmic}
\label{code:calculate_treat_effectiveness}
\end{algorithm}

\begin{table}[H]
\centering
\begin{tabular}{|>{\raggedright\arraybackslash}p{7cm}|>{\raggedright\arraybackslash}p{9cm}|}
\hline
\textbf{Name in Code} & \textbf{Description} \\ \hline
\texttt{treat\_patient()} & Simulates the treatment of a patient, calculates treatment effectiveness, and updates the doctor's experience and abilities based on the treatment outcome. This method also starts the treatment, marking the doctor as busy. \\ \hline
\texttt{update\_experience\_and\_abilities()} & Updates the doctor's experience level and possibly upgrades their credentials based on the accumulated experience and current research abilities. \\ \hline
\texttt{upgrade\_credential()} & Upgrades the doctor's credentials from low to medium or medium to high based on specific thresholds of research ability and experience. \\ \hline
\texttt{calculate\_treatment\_effectiveness()} & Calculates the effectiveness of the treatment based on the doctor's credentials, empathy, and technological resource constraints. The effectiveness is capped at 0.7. \\ 
\hline
\end{tabular}
\caption{Main Actions of the Doctor Agent in the Classical Model}
\label{tab:MethodsDoctorClassical}
\end{table}

\subsubsection{Patient Agent}

The patient agent is set up similarly to the doctor agent in the model. The patients in the model are heterogeneous; they have different health levels and both infected and unhealthy patients (patients who have low health level but are not infected) seek treatment from doctors when they are health level fall below $0.6$. The patient decides a doctor's desirability by taking into account the doctor's credential, mean rating from all the other patients who have been treated by the doctor and the patient's own past experience with the doctor. Based on an aggregate score of all the doctors in the simulation, the patient chooses a doctor if the doctor is not already attending to another patient and if the doctor was able to treat the patient satisfactorily (increasing the patient's health level to $0.8$ or above) in the past (assuming the past interaction had occurred). Subsequently, the patient rates the doctor out of a score of 5, with the perfect score being awarded to doctors who are able to increase the patients' health level to $0.8$ or beyond. The main properties that the patient has are as follows: 

\begin{enumerate}
    \item Health Level (\texttt{health\_level}): Health level is randomly initialized from $0.5$ and $1$ for each patient in the simulation. As mentioned previously, health level of a patient can be considered an aggregate score of medical measurements of how well vital organs of the patient are functioning. 

    \item Infection Resilience/Resistance to Treatment (\texttt{resilience}): This property, which is randomly initialized to take values between $0.1$ and $0.4$ for each patient, can be thought of as infection resistance to treatment and is a useful concept to understand how treatment effectiveness is modulated by random resistance to treatment in each patient \cite{Liu.etal2022a}. We use this treatment to scale the treatment effectiveness of the doctor for the patient's health level improvement. 

    \item Judgement weight for doctor's credential (\texttt{cred\_weight}): This is one of the weights for the patient's judgement score of the doctor, for the credential component of the judgement score. It is weighed roughly equally to the weight for the mean ratings from other patients in the system, and it is initiated randomly for each patient. We have included credential as one of the factors for a patient having a high regard for a doctor since it is one of the most abundantly available signals that a patient can find to evaluate the quality of the doctors, and it is considered to be one of the reliable signals in determining the quality of the doctor in the physical world \cite{Kozikowski.etal2022}. 

    \item Judgement weight for other patients' mean ratings for the doctor (\texttt{mean\_rating\_weight}): This is another weight for the patient's judgement of the doctor, for the mean rating from other patients. This weight can be thought of how much value a patient places on the "word of mouth" or public opinion regarding a doctor \cite{Harris2003}.

    \item Judgement weight for the patient's own past rating for the doctor (\texttt{past\_rating\_weight}): This is one of the most important weights for the judgement score, since it determines how much a patient values his own past rating regarding a doctor. This is given the highest weight range for the judgement score since it is the most reliable signal a patient can have regarding a doctor and has been found to be significant in a patient's decision in choosing a particular doctor \cite{Kozikowski.etal2022}.
\end{enumerate}

Table \ref{tab:patientpropertiesclassical} is given as a summary of all the main properties of the patient agent. 

\begin{table}[h!]
\centering
\begin{tabular}{|>{\raggedright\arraybackslash}p{5cm}|>{\raggedright\arraybackslash}p{8cm}|}
\hline
\textbf{Name in Code} & \textbf{Description} \\ \hline
\texttt{patient\_id} & Unique identifier for the patient. \\ \hline
\texttt{health\_level} & Current health level of the patient, initialized randomly between 0.5 and 1.0. \\ \hline
\texttt{resilience} & Patient's resilience factor, affecting recovery from illnesses, initialized randomly between 0.1 and 0.4. \\ \hline
\texttt{cred\_weight, mean\_rating\_weight, past\_rating\_weight} & Weights used by the patient to judge doctor performance, initialized randomly, summing to 1. \\ \hline
\end{tabular}
\caption{Main Properties of the Patient Agent in the Classical Model}
\label{tab:patientpropertiesclassical}
\end{table}

The main actions that a patient agent can enact in the simulation are:
\begin{enumerate}
    \item Patient Infection (\texttt{infect}): This method calculates the impact of an infection on the patient agent. If an agent is uninfected and has a health level of $0.1$ (ensuring that no patient agent dies in the simulation), the patient is infected and a value of $0.2$ is subtracted from his health level. Listing \ref{code:infect} and Equation \ref{equation:infect} are provided as an outline of how this action takes place in the simulation. 

    \begin{equation}
    H' = H - \chi{(\text{not infected} \land H > 0.1)} \times 0.2
    \label{equation:infect}
    \end{equation}
    \myequations{Patient Infection Process}
    where \( \chi \) is the indicator function that is 1 if the condition is true and 0 otherwise.
    
    \item Patient Priority (\texttt{priority}): The method keeps track of a list of infected patients and patients who have sufficiently low health in order to calculate a priority. Patients who are infected first gets the highest priority in terms of seeking treatment from a doctor and being able to choose the best available doctors in the simulation. This is in line with what takes place in the physical world, as patients who are infected first are often the ones who are able to choose from a greater number of doctors and take up space in the treatment queue \cite{Tracy.etal2018}. Listing \ref{code:priority} and Equation \ref{equation:priority} provide a more detailed look at how this method is implemented in the simulation. 

    \begin{equation}
    P = (I, \omega, H)
    \label{equation:priority}
    \end{equation}
    \myequations{Priority Logic}
    where \( I \) indicates infection status, \( \omega \) is the infection order or \( \infty \), and \( H \) is the health level.
    
    \item Doctor Requirement (\texttt{needs\_doctor}): A patient starts looking for a doctor when his health level is below $0.6$. We utilized a threshold model \cite{Djulbegovic.etal2014} for this method, and for the sake of simplicity we have implemented this to be the common value for all the patients in the simulation. Listing \ref{code:needs_doctor} depicts its implementation in our simulation. 

    \item Doctor Judgement (\texttt{judge\_doctor}): The method calculates a score for a doctor based on the doctor's credential, past rating of the patient and mean rating of other patients. Credentials are increasingly valued according to their categorical levels. Listing \ref{code:judge_doctor} and Equation \ref{equation:judge_doctor} summarizes the implementation in the model. The judgment score (\( J \)) for a doctor by a patient is calculated as:
    \begin{equation}
    J = w_c \cdot c + w_m \cdot m + w_p \cdot p
    \label{equation:judge_doctor}
    \end{equation}
    \myequations{Judging a Doctor}
    
    where:
    \begin{itemize}
        \item \( w_c, w_m, w_p \) are weights for the doctor's credentials, mean rating from all patients, and past rating by this patient, respectively.
        \item \( c \) is the score assigned based on the doctor's credential: \{low: 0.1, medium: 0.5, high: 1.0\}.
        \item \( m \) is the mean rating of the doctor across all patients.
        \item \( p \) is the past rating given by this patient, if any.
    \end{itemize}

    \item Choosing a Doctor (\texttt{choose\_doctor}): The method enables the patient to select a doctor based on availability, past ratings and judgements. It first filters out doctors who are currently busy. If the patient has previously received a perfect rating of $5$ from a specific doctor and that doctor is available, he continues with that same doctor. If not, the method evaluates the remaining doctors based on the judgment score. The doctor with the highest judgment score is then chosen, unless no doctors meet the criteria, in which case no doctor is selected. This process ensures that patients are matched with doctors who are best suited to their needs based on past interactions and overall performance ratings. Listing \ref{code:choose_doctor} and Equation \ref{equation:choose_doctor} highlight the implementation of the method, with additional methods and variables used in the code for simulation purposes. 

    \begin{equation}
    D^* = \arg \max_{D \in \mathcal{D}} J_D
    \label{equation:choose_doctor}
    \myequations{Choosing a Doctor}
    \end{equation}
    where \( J_D \) is the judgment score of doctor \( D \), calculated as previously defined. If a doctor is busy or not satisfactory based on past treatments, she is excluded from \( \mathcal{D} \).

    \item Receiving Treatment (\texttt{receive\_treatment}): It facilitates the treatment process where a patient receives medical care from a selected doctor. The method calculates the treatment's effectiveness by considering both the doctor's skills and the patient's resilience to determine the actual impact on the patient's health. After treatment, the patient's health level is updated based on this calculated effectiveness, and the patient's infection status is reset to indicate they are no longer infected. Additionally, the patient rates the doctor based on the effectiveness of the treatment, contributing to the doctor's overall performance evaluation. This method encapsulates the interaction between patient health management and feedback mechanisms within the healthcare simulation. The method is depicted in Listing \ref{code:receive_treatment} and Equation \ref{equation:receive_treatment} with more specific details of its implementation in the simulation. 

    \begin{equation}
    H = \max(0.1, \min(1, H + E \cdot (1 - R)))
    \label{equation:receive_treatment}
    \myequations{Receiving Treatment Process}
    \end{equation}
    where \( E \) is treatment effectiveness and \( R \) is resilience.

    \item Update Health Level (\texttt{update\_health\_level}): This method is responsible for adjusting a patient's health level after receiving treatment. It uses the effectiveness of the treatment to increase the health level, ensuring that the final value remains within the bounds of $0.1$ (minimum) and $1.0$ (maximum). This update reflects the impact of medical intervention on the patient's health, contributing to a dynamic simulation of patient health over time. The method also records this updated health level in the patient's health history, allowing for tracking and analysis of health trends throughout the simulation. The implementation is outlined in Listing \ref{code:update_health_level} .

    \item Rating a Doctor (\texttt{rate\_doctor}): After receiving treatment, the patient rates the doctor based on the improved health level. If the patient’s health level reaches or exceeds a threshold ($0.8$ in this context), the doctor receives the highest rating ($5$). If not, the rating is calculated proportionally to how close the health level is to this threshold. Once again, much like the patient's search behaviour, we utilize a threshold-based approach to the rating system. This rating is then recorded in the rating system, which tracks and aggregates doctor performances over time. This method plays a crucial role in providing feedback on a doctor's ability and effectiveness, influencing future patient choices and doctor improvements in the simulation. The implementation is outlined in Listing \ref{code:rate_doctor}  and Equation \ref{equation:rating}.
    
    \begin{equation}
    R = \begin{cases} 
    5 & \text{if } h \geq 0.8 \\
    \max(0, \text{int}(5 \cdot \frac{h}{0.8})) & \text{otherwise}
    \end{cases}
    \label{equation:rating}
    \myequations{Rating Proccess}
    \end{equation}
    
    where \( h \) is the health level of the patient after receiving treatment.
\end{enumerate}

Table \ref{tab:methods_patient_classical} is given as a summary of all the actions a doctor agent can take in the model. 

\begin{table}[H]
\centering
\begin{tabular}{|l|p{10cm}|}
\hline
\textbf{Name in Code} & \textbf{Description} \\ \hline
\texttt{infect()} & Infects the patient if they are not already infected and their health level is above 0.1, decreasing their health by a set severity level and marking them as infected. \\ \hline
\texttt{priority()} & Returns a tuple representing the patient's priority for treatment, based on their infection status, order of infection, and health level. \\ \hline
\texttt{needs\_doctor()} & Checks if the patient's health level is below 0.6, indicating the need for a doctor. \\ \hline
\texttt{judge\_doctor()} & Computes a judgment score for a doctor based on the doctor's credentials, average patient ratings, and personal past interactions with the patient. \\ \hline
\texttt{choose\_doctor()} & Selects a doctor from a list of available doctors based on their judgment scores and previous interactions with the patient. \\ \hline
\texttt{receive\_treatment()} & Facilitates the process where the patient receives treatment from a doctor, updates their health level, and provides a rating for the doctor. \\ \hline
\texttt{update\_health\_level()} & Updates the patient's health level based on the effectiveness of the treatment received. \\ \hline
\texttt{rate\_doctor()} & Provides a rating for the doctor based on the effectiveness of the treatment, adjusting the doctor's rating in the rating manager. \\ \hline
\end{tabular}
\caption{Main Actions of the Patient Agent in the Classical Model}
\label{tab:methods_patient_classical}
\end{table}

\begin{algorithm}[h!]
\caption{Infect}
\begin{algorithmic}[1]
\Function{Infect}{order}
    \If{not self.is\_infected \textbf{and} self.health\_level $>$ 0.1}
        \State self.health\_level $\gets$ max(0, self.health\_level - 0.2)
        \State self.is\_infected $\gets$ \textbf{true}
        \State self.infected\_order $\gets$ order
        \State \textbf{print} "Patient \{self.patient\_id\} has been infected and is number \{order\} in line. Health level reduced to \{self.health\_level:.2f\}"
    \EndIf
\EndFunction
\end{algorithmic}
\label{code:infect}
\end{algorithm}

\begin{algorithm}[h!]
\caption{Priority}
\begin{algorithmic}[1]
\Function{Priority}{}
    \If{self.is\_infected}
        \State \Return (False, self.infected\_order, self.health\_level)
    \Else
        \State \Return (True, $\infty$, self.health\_level)
    \EndIf
\EndFunction
\end{algorithmic}
\label{code:priority}
\end{algorithm}

\begin{algorithm}[h!]
\caption{Needs Doctor}
\begin{algorithmic}[1]
\Function{NeedsDoctor}{}
    \If{self.health\_level $<$ 0.6}
        \State \Return \textbf{true}
    \Else
        \State \Return \textbf{false}
    \EndIf
\EndFunction
\end{algorithmic}
\label{code:needs_doctor}
\end{algorithm}

\begin{algorithm}[H]
\caption{Choose a Doctor}
\begin{algorithmic}[1]
\Function{choose\_doctor}{doctors}
    \If{self.needs\_doctor()}
        \State available\_doctors $\gets$ Empty List
        \For{doc in doctors}
            \If{not doc.is\_busy}
                \State Append doc to available\_doctors
            \EndIf
        \EndFor
        \If{available\_doctors is empty}
            \State Print ``No available doctors for patient self.patient\_id at this time.''
            \State \Return None
        \EndIf
        \For{doc in available\_doctors}
            \If{self.last\_doctor\_id \textbf{and} self.rating\_manager.get\_rating\_by\_patient(self.last\_doctor\_id, self.patient\_id) == 5}
                \State \Return doc
            \EndIf
        \EndFor
        \State eligible\_doctors $\gets$ Empty List
        \For{doc in available\_doctors}
            \If{doc.get\_id() $\neq$ self.last\_doctor\_id or self.rating\_manager.get\_rating\_by\_patient(doc.get\_id(), self.patient\_id) $\neq$ 5}
                \State Append doc to eligible\_doctors
            \EndIf
        \EndFor
        \If{eligible\_doctors}
            \State \Return doc with highest self.judge\_doctor(doc) among eligible\_doctors
        \Else
            \State \Return doc with highest self.judge\_doctor(doc) among available\_doctors
        \EndIf
    \EndIf
    \State \Return None
\EndFunction
\end{algorithmic}
\label{code:choose_doctor}
\end{algorithm}

\begin{algorithm}[h!]
\caption{Receive Treatment}
\begin{algorithmic}[1]
\Function{ReceiveTreatment}{doctor}
    \State effectiveness $\gets$ doctor.treat\_patient() * (1 - self.resilience)
    \State \Call{UpdateHealthLevel}{effectiveness}
    \State self.is\_infected $\gets$ \textbf{false}
    \State rating $\gets$ \Call{RateDoctor}{doctor}
    \State \textbf{print} "Patient \{self.patient\_id\} treated by Doctor \{doctor.get\_id()\}, rating: \{rating\}. Health: \{self.health\_level:.2f\}"
\EndFunction
\end{algorithmic}
\label{code:receive_treatment}
\end{algorithm}

\begin{algorithm}[h!]
\caption{Update Health Level}
\begin{algorithmic}[1]
\Function{UpdateHealthLevel}{effectiveness}
    \State self.health\_level $\gets$ \textbf{max}(0.1, \textbf{min}(1, self.health\_level + effectiveness))
    \State \Call{AppendHealthHistory}{self.health\_level}
\EndFunction
\end{algorithmic}
\label{code:update_health_level}
\end{algorithm}

\begin{algorithm}[h!]
\caption{Rate a Doctor}
\begin{algorithmic}[1]
\Function{rate\_doctor}{doctor}
    \If{self.health\_level $\geq$ 0.8}
        \State $\text{rating} \gets 5$
    \Else
        \State $\text{rating} \gets \max(0, \text{int}(5 \times (\text{self.health\_level} / 0.8)))$
    \EndIf
    \State $\text{self.last\_doctor\_id} \gets \text{doctor.get\_id()}$
    \State \Return $\text{rating}$
\EndFunction
\end{algorithmic}
\label{code:rate_doctor}
\end{algorithm}

\begin{algorithm}[H]
\caption{Judge a Doctor}
\begin{algorithmic}[1]
\Function{judge\_doctor}{doctor}
    \State $\text{credential\_score} \gets \{\text{'low'}: 0.1, \text{'medium'}: 0.5, \text{'high'}: 1.0\}[\text{doctor.get\_credential}()]$
    \State $\text{mean\_rating} \gets \text{self.rating\_manager.get\_mean\_rating}(\text{doctor.get\_id}())$
    \State $\text{past\_rating} \gets \text{self.rating\_manager.get\_rating\_by\_patient}(\text{doctor.get\_id}(), \text{self.patient\_id}) \text{ or } 0$
    \State $\text{judgment} \gets \text{self.cred\_weight} \times \text{credential\_score} + \text{self.mean\_rating\_weight} \times \text{mean\_rating} + \text{self.past\_rating\_weight} \times \text{past\_rating}$
    \State \Return $\text{judgment}$
\EndFunction
\end{algorithmic}
\label{code:judge_doctor}
\end{algorithm}

\subsection{Cognitive Social System}

The cognitive social system model is similar to the classical model, with a few, key extensions. The main difference between the models is that cognitive social system integrates ideas from reflexivity and cognitive social structures. These ideas are incorporated through both the doctor and the patient agent. Now both the doctor and the patient agent in the model possess social strength values for all the agents, apart from herself or himself, present in the model, with values ranging from $0$ to $1$; $S \in [0,1]$ where $S$ is denoted as the strength of the social tie an agent has with all the other agents in the simulation. 

\begin{figure}[h!]
    \centering
    \includegraphics[width=1\linewidth]{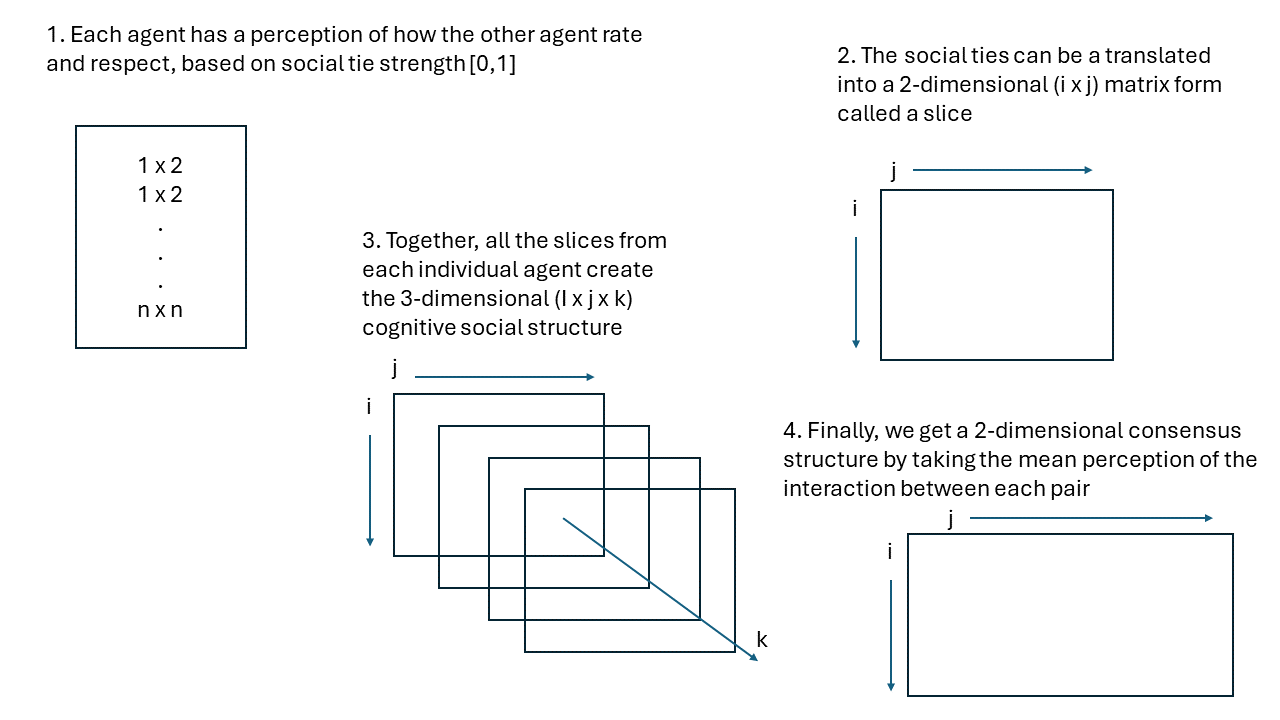}
    \caption{Social Matrix as conceived in Cognitive Social Structure.}
    \label{fig:CSS_Outline}
\end{figure}

The main idea of cognitive social structures is that there is a relationship $R_{i,j,k}$, which represents the perception of how $i$ interacts with $j$ through the lens of $k$ \cite{Brands2013}. The way other patients rate a doctor is modulated by the social ties the other patients have with the doctors, so the value of their ratings are also weighed by their respective social strength to the patient who is judging a doctor to select and receive treatment from. This behaviour is summarized in Figure \ref{fig:CSS_Outline}, which has been adapted from a paper by Frank (2015), where the author applied cognitive social structure analysis to investigate the group social norms around around H1N1 flu prevention \cite{Frank2015}. In this model, we utilize cognitive social structures to explore how patients judge and choose doctors based on an aggregate consensus, mean weighted rating, of the doctor's abilities for infection treatments.

\begin{figure}[h!]
    \centering
    \includegraphics[width=1\linewidth]{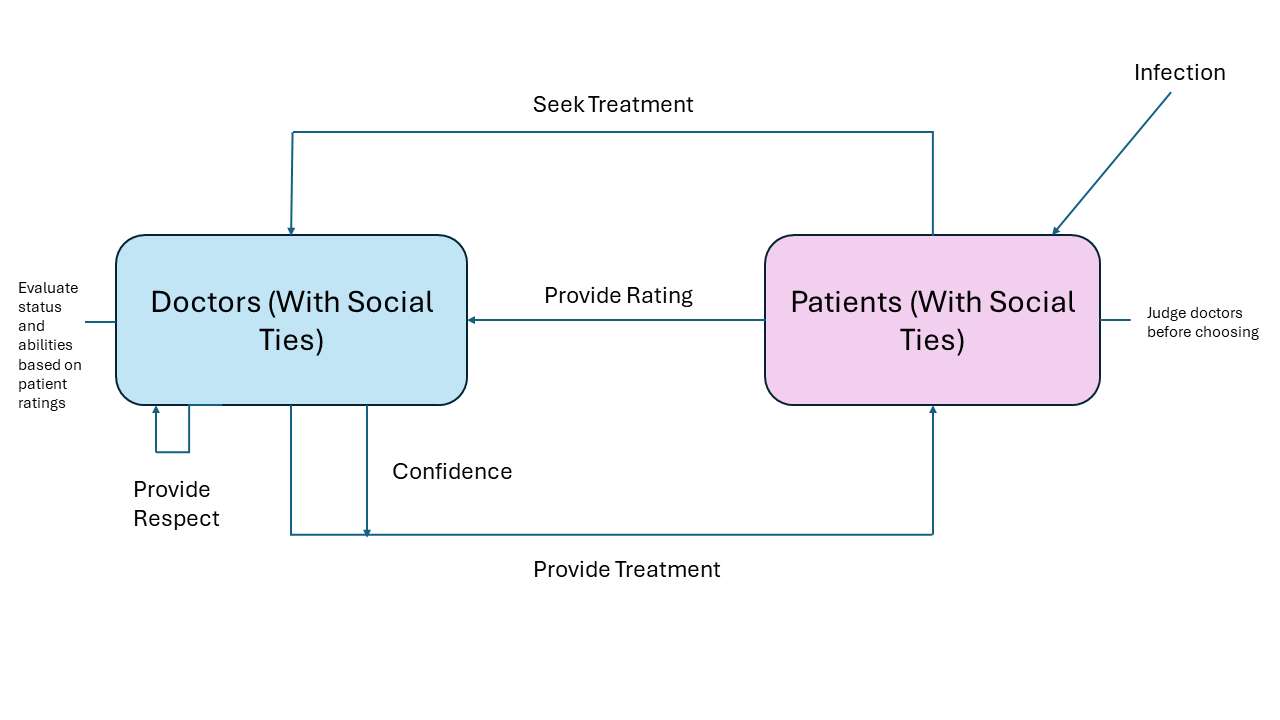}
    \caption{Cognitive Social System Agent Interacttions}
    \label{fig:CSS_Interactions}
\end{figure}

The doctors and patients have the same objectives as before. The patients receive feedback for their choices of doctors by noting how much of their health has been recovered from the treatment, but now it is modulated by the social strength that each patient has for a specific doctor. The doctors receive feedback for their treatment through ratings by patients which is dependent on how well the doctors' treatments were able to help patients in recovering their health level, which is once again modulated by social strengths. The doctors now also have respect values for each other to calculate confidence, which is our way of implementing positive feedback loop between doctors to fuse reflexivity into the model. Figure \ref{fig:CSS_Interactions} summarizes the main interaction logic for the agents in the classical model. 

\subsubsection{Doctor Agent}

Apart from social tie values for both doctors and patients being added as properties, there are three new properties that the doctor agent has in this model that are different from the previous model. They are used for calculating the respect values for one's colleagues and ultimately, one's confidence which affect the treatment effectiveness. They are as follows; 

\begin{enumerate}
    \item Respect for Colleagues (\texttt{respect\_for\_colleagues}): This property is a way to capture a cognitive aspect of the doctors, namely how respect for each other can create a form of social circle which can either deflate or encourage doctors to improve their treatment method and therefore, treatment effectiveness \cite{Lipworth.etal2013}. It is initially set to $0$, but it changes based on the interaction with the doctors. 

    \item Confidence (\texttt{confidence}): Confidence values are initially set to $0$, and they change depending on the ratings received from patients, whose effects are modulated by social tie strengths, and perceived respect derived from other doctors. 

    \item Weight for Mean Weighted Ratings from Patients (\texttt{weight\_wmrat}): This is one of the weights that is used to calculate the value of confidence derived from the ratings. The underlying logic for it is that higher the ratings, the confidence values should proportionately be higher. 

    \item Weight for Mean Weighted Respects from Colleagues (\texttt{weight\_mres}): The second weight for the confidence calculation. The logic is similar to the other weight; if colleagues respect you more, you are more likely to approach a treatment more confidently and with greater chances of success.
\end{enumerate}

Table \ref{tab:doctor_attributes_cognitive} outlines the key changes in properties for the doctor agent.

\begin{table}[h]
\centering
\begin{tabular}{|l|p{10cm}|}
\hline
\textbf{Name in Code} & \textbf{Description} \\
\hline
social\_ties\_doctors & Dictionary representing the strength of social ties to other doctors. \\
\hline
social\_ties\_patients & Dictionary representing the strength of social ties to patients. \\
\hline
respect\_for\_colleagues & Initially zero for each doctor, can change based on interactions. \\
\hline
confidence & A measure of the doctor's self-confidence, influenced by ratings and respects received. \\
\hline
weight\_wmrat & Weight for mean weighted ratings from patients. \\
\hline
weight\_mwres & Weight for mean weighted respects from colleagues. \\
\hline
\end{tabular}
\caption{Additional Properties of the Doctor Class in the Cognitive Social System}
\label{tab:doctor_attributes_cognitive}
\end{table}

There are some additional actions a doctor undertakes that can lead to changes in calculations for treatment effectiveness in this model. We will provide the new actions for the doctor agent below;

\begin{enumerate}
    \item Calculate Mean Weighted Respects (\texttt{calculate\_mean\_weighted\_respects}): This method calculates the average respect that a doctor receives from their colleagues, factoring in the strength of social ties between them. It iterates through all other doctors in the system, gathering the respect values directed towards the doctor in question. These respect values are then weighted by the strength of the social ties (how well the doctors know each other or their professional closeness), and an average is computed based on these weighted values. Listing \ref{code:mean_weighted_respects} and Equation \ref{equation:mean_weighted_respects} provide detailed technical overview regarding its implementation. 

    \begin{equation}
    \text{MeanWeightedRespects} = \frac{\sum_{i \neq \text{self}} R_i \cdot S_i}{\sum_{i \neq \text{self}} S_i} 
    \label{equation:mean_weighted_respects}
    \myequations{Mean Weighted Respect Calculation}
    \end{equation}
    \text{where } $R_i$ \text{ is the respect from doctor } i, \text{ and } $S_i$ \text{ is the social strength with doctor } i.

    \item Respect for Colleagues (\texttt{update\_respect\_for\_colleagues}): The method assesses each colleague's credentials and the ratings they've received from patients. These assessments are then adjusted by the strength of the social ties the evaluating doctor has with the patients who have rated the colleague. The resulting values update the doctor’s respect levels for each colleague. Listing \ref{code:calculate_respect_for_colleagues} and Equation \ref{equation:update_respect_colleague} outline the key aspects of this method.

    \begin{equation}
    \text{NewRespect}_{j} = S_{j} \cdot (C_{j} + \text{Weighted\_Valutation}_{j}) 
    \label{equation:update_respect_colleague}
    \myequations{Update Respect Logic}
    \end{equation}
    where $S_{j}$ is the social strength to $doctor_j$, $C_{j}$ is the credential score for $doctor_j$, and ${WeightedRating}_{j}$  is the weighted valuation of ratings received by $doctor_j$.

    The weighted valuation of a doctor's ratings based on another doctor's social ties can be calculated using the following equation:

    \begin{equation}
    \text{Weighted\_Valuation} = \sum_{i=1}^{n} r_i \times s_{i}
    \label{equation:weighted_valuation}
    \myequations{Weighted Valuation Equation}
    \end{equation}
    
    where:
    \begin{itemize}
        \item \( r_i \) represents the rating given by the \( i \)-th patient,
        \item \( s_i \) is the strength of the social tie between the evaluating doctor and the \( i \)-th patient,
        \item \( n \) is the total number of patients who have rated the evaluated doctor.
    \end{itemize}
    
    This formula assumes that each rating \( r_i \) is weighted by \( s_i \), the strength of the social tie to the evaluating doctor. If there is no social tie between the evaluating doctor and a patient (\( s_i = 0 \)), that patient's rating does not contribute to the weighted valuation.

    \item Confidence Calculation (\texttt{update\_confidence}): This method computes the doctor's confidence by combining two key elements: the mean weighted ratings from patients and the mean weighted respects from colleagues. These elements are each given a certain weight (importance), and the resulting weighted sum defines the new confidence level. Listing \ref{code:confidence_calc} and Equation \ref{equation:confidence_calculation} summarize the process. 

    \begin{equation}
    \text{Confidence} = W_{\text{ratings}} \cdot \text{MeanWeightedRatings} + W_{\text{respects}} \cdot \text{MeanWeightedRespects} 
    \label{equation:confidence_calculation}
    \myequations{Confidence Equation}
    \end{equation}
    where $W_{ratings}$ and $W_{respects}$ are the weights assigned to the mean weighted ratings and respects, respectively.
    
    The mean weighted ratings for a doctor based on the social ties to patients can be calculated using the following formula:

    \begin{equation}
    \text{Mean Weighted Rating} = 
    \begin{cases} 
    \frac{\sum_{i=1}^n r_i \cdot s_i}{\sum_{i=1}^n s_i} & \text{if } \sum_{i=1}^n s_i > 0 \\
    0 & \text{otherwise}
    \end{cases}
    \myequations{Mean Weighted Rating}
    \label{equation:mean_weighted_rating}
    \end{equation}
    
    where:
    \begin{itemize}
        \item \( r_i \) represents the rating given by the \( i \)-th patient,
        \item \( s_i \) is the strength of the social tie between the doctor and the \( i \)-th patient,
        \item \( n \) is the number of patients who have rated the doctor.
    \end{itemize}

    If the total strength of all social ties (\( \sum_{i=1}^n s_i \)) is greater than zero, the mean weighted rating is calculated by dividing the sum of all weighted ratings (\( \sum_{i=1}^n r_i \cdot s_i \)) by the sum of all social strengths. If the total social strength is zero, the mean weighted rating is defined to be zero.

    \item Treatment Effectiveness Calculation (\texttt{calculate\_treatment\_effectiveness}): It assesses the effectiveness of a doctor’s treatment based on several key factors. It integrates the doctor's credentials, empathy and any external technological resource constraints to calculate the overall treatment effectiveness, as before. However, now it also includes confidence in calculating the effectiveness. The result reflects the combined impact of the doctor’s professional competence, interpersonal skills, self-assuredness and any limitations imposed by their work environment, with the final value being capped to ensure it stays within realistic bounds. Listing \ref{code:cognitive_treatment_effect} and Equation \ref{equation:effectiveness_cognitive} are provided to illustrate the new calculation. 
    
    \begin{equation}
    \text{Effectiveness} = \min \left( (C_{\text{self}} + E_{\text{self}} + \text{CD}) \cdot (1 - T_{\text{self}}), 0.7 \right) 
    \label{equation:effectiveness_cognitive}
    \myequations{Treatment Effectiveness with Social Ties}
    \end{equation}
    where $C_{self}$ is the credential factor, $E_{\text{self}}$ is the empathy level, \text{CD} is the current confidence level, and  $T_{\text{self}}$  is the technological resource constraint.
\end{enumerate}

\begin{table}[H]
\centering
\begin{tabular}{|l|p{10cm}|}
\hline
\textbf{Name in Code} & \textbf{Description} \\
\hline
calculate\_treatment\_effectiveness & Determines how effective the doctor's treatment is, based on their credentials, empathy, and technological resources. \\
\hline
calculate\_mean\_weighted\_respects & Computes the average respect received from colleagues, weighted by social ties. \\
\hline
update\_respect\_for\_colleagues & Updates the respect scores for colleagues based on their credentials and patient ratings. \\
\hline
update\_confidence & Updates the doctor's confidence level based on weighted patient ratings and respects from colleagues. \\
\hline
\end{tabular}
\caption{Additional Actions of the Doctor Agent in the Cognitive Social System}
\label{tab:doctor_actions_cognitive}
\end{table}

Table \ref{tab:doctor_actions_cognitive} highlights the key updates in actions for the doctor agent. The actions of the doctor agent utilizes key calculations from rating manager class in our simulation, \texttt{calculate\_mean\_weighted\_ratings} (Listing \ref{code:calculuate_mean_weighted_ratings} and Equation \ref{equation:mean_weighted_rating}) and \texttt{calculate\_weighted\_valuation} (Listing \ref{code:weighted_valuation} and \ref{equation:weighted_valuation}).

\begin{algorithm}[H]
\caption{Calculate Mean Weighted Respects}
\begin{algorithmic}[1]
\Procedure{CalculateMeanWeightedRespects}{all\_doctors}
    \State $total\_weighted\_respects \gets 0$
    \State $total\_social\_strengths \gets 0$
    \For{each doctor in all\_doctors}
        \If{doctor.doctor\_id $\neq$ self.doctor\_id}
            \State $respect \gets doctor.respect\_for\_colleagues[self.doctor\_id]$
            \State $social\_strength \gets self.social\_ties\_doctors[doctor.doctor\_id]$
            \State $total\_weighted\_respects \gets total\_weighted\_respects + (respect \times social\_strength)$
            \State $total\_social\_strengths \gets total\_social\_strengths + social\_strength$
        \EndIf
    \EndFor
    \If{$total\_social\_strengths > 0$}
        \State \Return $total\_weighted\_respects / total\_social\_strengths$
    \Else
        \State \Return $0$
    \EndIf
\EndProcedure
\end{algorithmic}
\label{code:mean_weighted_respects}
\end{algorithm}

\begin{algorithm}
\caption{Respect for Colleagues}
\begin{algorithmic}[1]
\Procedure{UpdateRespectForColleagues}{all\_doctors, rating\_manager}
    \For{each other\_doctor in all\_doctors}
        \If{other\_doctor.doctor\_id $\neq$ self.doctor\_id}
            \State $weighted\_valuation \gets rating\_manager.get\_weighted\_valuation(other\_doctor.doctor\_id,$ 
            \State $self.social\_ties\_patients)$
            \State $credential\_score \gets credential\_factor[other\_doctor.credential]$
            \State $total\_valuation \gets credential\_score + weighted\_valuation$
            \State $social\_strength \gets self.social\_ties\_doctors[other\_doctor.doctor\_id]$
            \State $self.respect\_for\_colleagues[other\_doctor.doctor\_id] \gets social\_strength \times total\_valuation$
        \EndIf
    \EndFor
\EndProcedure
\end{algorithmic}
\label{code:calculate_respect_for_colleagues}
\end{algorithm}

\begin{algorithm}
\caption{Confidence Calculation}
\begin{algorithmic}[1]
\Procedure{UpdateConfidence}{rating\_manager, all\_doctors}
    \State $mean\_weighted\_ratings \gets rating\_manager.calculate\_mean\_weighted\_ratings(self.doctor\_id,$ 
    \State $self.social\_ties\_patients)$
    \State $mean\_weighted\_respects \gets self.calculate\_mean\_weighted\_respects(all\_doctors)$
    \State $self.confidence \gets (self.weight\_wmrat \times mean\_weighted\_ratings) + (self.weight\_mwres \times mean\_weighted\_respects)$
\EndProcedure
\end{algorithmic}
\label{code:confidence_calc}
\end{algorithm}

\begin{algorithm}[H]
\caption{Treatment Effectiveness Calculation}
\begin{algorithmic}[1]
\Procedure{CalculateTreatmentEffectiveness}{}
    \State $credential\_factor \gets \{'low': 0.1, 'medium': 0.2, 'high': 0.3\}$
    \State $effectiveness \gets (credential\_factor[self.credential] + self.empathy + self.confidence) \times (1 - self.technological\_resource\_constraint)$
    \State \Return $\min(effectiveness, 0.7)$
\EndProcedure
\end{algorithmic}
\label{code:cognitive_treatment_effect}
\end{algorithm}

\begin{algorithm}[H]
\caption{Calculate Mean Weighted Ratings Based on Social Ties}
\begin{algorithmic}[1]

\Function{CalculateMeanWeightedRatings}{doctor\_id, social\_ties\_patients}
    \If{doctor\_id in self.ratings \textbf{and} self.ratings[doctor\_id]}
        \State $total\_weighted\_rating \gets 0$
        \State $total\_social\_strengths \gets 0$
        \For{(patient\_id, rating) in self.ratings[doctor\_id].items()}
            \State $social\_strength \gets$ social\_ties\_patients.get(patient\_id, 0)
            \If{$social\_strength > 0$}
                \State $total\_weighted\_rating \gets total\_weighted\_rating + rating \times social\_strength$
                \State $total\_social\_strengths \gets total\_social\_strengths + social\_strength$
            \EndIf
        \EndFor
        \If{$total\_social\_strengths > 0$}
            \State \Return $total\_weighted\_rating / total\_social\_strengths$
        \Else
            \State \Return 0
        \EndIf
    \Else
        \State \Return 0
    \EndIf
\EndFunction

\end{algorithmic}
\label{code:calculuate_mean_weighted_ratings}
\end{algorithm}

\begin{algorithm}[H]
\caption{Calculate Weighted Valuation Based on Social Ties}
\begin{algorithmic}[1]

\Function{GetWeightedValuation}{doctor\_evaluated\_id, social\_ties\_from\_evaluator}
    \If{doctor\_evaluated\_id in self.ratings \textbf{and} self.ratings[doctor\_evaluated\_id]}
        \State $total\_weighted\_rating \gets 0$
        \For{(patient\_id, rating) in self.ratings[doctor\_evaluated\_id].items()}
            \State $social\_tie\_strength \gets$ social\_ties\_from\_evaluator.get(patient\_id, 0)
            \State $total\_weighted\_rating \gets total\_weighted\_rating + rating \times social\_tie\_strength$
        \EndFor
        \State \Return $total\_weighted\_rating$
    \Else
        \State \Return 0
    \EndIf
\EndFunction

\end{algorithmic}
\label{code:weighted_valuation}
\end{algorithm}

\subsubsection{Patient Agent}

The patient agent has no new properties except for social ties which are similar to the doctor agent; each doctor and each patient is assigned a random value between $0$ and $1$ to indicate the strength of the social tie to the patient agent in question. However, how a patient judges a doctor and rates a doctor have now changed. The description of the changes are provided below; 

\begin{enumerate}
    \item Doctor Judgement (\texttt{judge\_doctor}): The method evaluates a doctor based on several weighted factors, including the doctor's credentials, mean ratings from other patients, and the patient's own past experiences with the doctor, which is the same as before. However, now, each factor is weighted by the patient's social ties to the doctor and other patients who have rated the doctor. Pseudocode \ref{code:judge_doctor_cognitive} and Equation \ref{equation:judge_doctor_cognitive} are given to depict the update due to the model. 

    \begin{equation}
    \text{Judgment} = W_{\text{cred}} \cdot (C \cdot S_{\text{doc}}) + W_{\text{mean\_rat}} \cdot \left(\frac{\sum (R_{p} \cdot S_{p})}{\sum S_{p}}\right) + W_{\text{past\_rat}} \cdot R_{\text{past}}
    \label{equation:judge_doctor_cognitive}
    \end{equation}
    \text{where:}
    \begin{conditions}
     W_{\text{cred}}, W_{\text{mean\_rat}}, W_{\text{past\_rat}} & \text{weights for credential, mean ratings, and past rating} \\
     C & \text{credential score of the doctor} \\
     S_{\text{doc}} & \text{social tie strength with the doctor} \\
     R_{p} & \text{rating from other patients} \\
     S_{p} & \text{social tie strength with other patients who have rated the doctor} \\
     R_{\text{past}} & \text{past rating given by this patient to the doctor}
    \end{conditions}
    \myequations{Judgment Calculation with Social Ties}

    \item Rating a Doctor (\texttt{rate\_doctor}): The implementation of this method is more or less the same as implemented in the previous model. But now the final rating now is adjusted based on the social strength, emphasizing how the interpersonal relationship can influence the patient's perceived quality of care. Pseudocode \ref{code:rate_doctor_cognitive} and Equation \ref{equation:rate_doctor_cogntive} are once again provided to note the differences. 

\begin{equation}
    \text{Rating} = \min\left(5, \text{round}\left( \text{BaseRating} \cdot (1 + 0.1 \cdot S_{d}), 1 \right) \right)
    \label{equation:rate_doctor_cogntive}
\end{equation}
\text{where:}
\begin{align}
    \text{BaseRating} & = 
    \begin{cases} 
    5 & \text{if } H \geq 0.8 \\
    \max\left(0.0, 5.0 \cdot \left(\frac{H}{0.8}\right)\right) & \text{otherwise}
    \end{cases} \\
    H & : \text{health level of the patient after treatment} \\
    S_{d} & : \text{social tie strength with the doctor}
\end{align}

\end{enumerate}

\begin{algorithm}[H]
\caption{Judge Doctor with Social Strength}
\begin{algorithmic}[1]
\Procedure{JudgeDoctor}{doctor}
    \State $credential\_score \gets \{'low': 0.1, 'medium': 0.5, 'high': 1.0\}[doctor.get\_credential()]$
    \State $weighted\_credential\_score \gets self.social\_ties\_doctors[doctor.get\_id()] \cdot credential\_score$
    \State $total\_weighted\_ratings \gets 0$
    \State $total\_social\_strength \gets 0$
    \For{each patient\_id, social\_strength in self.social\_ties\_patients}
        \State $patient\_rating \gets self.rating\_manager.get\_rating\_by\_patient(doctor.get\_id(), patient\_id)$
        \If{$patient\_rating \neq \text{None}$}
            \State $total\_weighted\_ratings \gets total\_weighted\_ratings + patient\_rating \cdot social\_strength$
            \State $total\_social\_strength \gets total\_social\_strength + social\_strength$
        \EndIf
    \EndFor
    \State $mean\_weighted\_ratings \gets total\_weighted\_ratings / total\_social\_strength$ \textbf{if} $total\_social\_strength > 0$ \textbf{else} $0$
    \State $past\_rating \gets self.rating\_manager.get\_rating\_by\_patient(doctor.get\_id(), self.patient\_id) \textbf{or} 0$
    \State $judgment \gets self.cred\_weight \cdot weighted\_credential\_score + self.mean\_rating\_weight \cdot mean\_weighted\_ratings + self.past\_rating\_weight \cdot past\_rating$
    \State \Return $judgment$
\EndProcedure
\end{algorithmic}
\label{code:judge_doctor_cognitive}
\end{algorithm}

\begin{algorithm}[H]
\caption{Rate Doctor with Social Strength}
\begin{algorithmic}[1]
\Procedure{RateDoctor}{doctor}
    \State $effectiveness \gets doctor.treat\_patient(all\_doctors) \cdot (1 - self.resilience)$
    \State $self.update\_health\_level(effectiveness)$
    \State $self.treat\_infection()$
    \If{$self.health\_level \geq 0.8$}
        \State $base\_rating \gets 5$
    \Else
        \State $base\_rating \gets \max(0.0, (5.0 \cdot (self.health\_level / 0.8)))$
    \EndIf
    \State $social\_strength \gets self.social\_ties\_doctors[doctor.get\_id()]$
    \State $adjusted\_rating \gets base\_rating \cdot (1 + 0.1 \cdot social\_strength)$
    \State $rating \gets \min(5.0, \text{round}(adjusted\_rating, 1))$
    \State $self.rating\_manager.add\_rating(doctor.get\_id(), self.patient\_id, rating)$
    \State \Return $rating$
\EndProcedure
\end{algorithmic}
\label{code:rate_doctor_cognitive}
\end{algorithm}

\subsection{Micorbial Genetic Algorithm}

A genetic algorithm, first introduced by Holland (1984), is a heuristic algorithm that is designed to reflect the process of natural selection where the fittest individuals are selected for reproduction to produce offspring of the next generation \cite{Holland1984}. It simulates processes in natural systems necessary for evolution, particularly those that follow the principles of survival of the fittest to reproduce the next generation \cite{Holland1984}. The key components of a genetic algorithm include; 

\begin{enumerate}
    \item Population: A set of solutions for a given problem; each solution or an individual is encoded with a set of parameters also known as chromosomes.

    \item Fitness Function: A function that evaluates and ranks individuals according to how well they solve the problem. 
    
    \item Selection: A method to select fittest individuals to pass their genes or traits to the next generation. It can involve techniques like roulette wheel selection and tournament selection among others. 
    
    \item Crossover (Recombination): A process to combine genetic information of two parents to generate new offspring. 
    
    \item Mutation: A way to introduce genetic variation in the population by randomly changing individual genes or traits of offspring. It prevents the algorithm from getting stuck in a local optima by maintaining a genetic diversity within the population. 
\end{enumerate}

For our research, we chose to implement a more streamlined version of the genetic algorithm known as microbial genetic algorithm. Our decision to implement a genetic algorithm rests on the fact that agents such as doctor and patient can be thought of as having traits or characteristics which can change after interactions as the agents learn from their encounters. The best agents are the ones that are able to fulfil their objectives successfully and therefore are likely to possess traits that represent abilities of the agents that enable them to reach their objectives \cite{Farago.etal2022}. 

A microbial genetic algorithm is a simplified variant of the traditional genetic algorithm outlined above, characterized by its more direct approach to simulating the evolutionary process \cite{Harvey2011}. The algorithm operates on a smaller population and specifically uses a tournament-style selection process; two individuals are chosen at random to compete and the "loser" is replaced by the mutated version of the " winner", mimicking a microbial infection spreading through a population \cite{Harvey2011}. We implemented a microbial genetic algorithm instead of a conventional one because the microbial algorithm is efficient in its convergence, and our simulation of doctor-patient interaction is relatively simple for the algorithm to effectively maintain genetic diversity within the population. Key components of the algorithm include; 

\begin{enumerate}
    \item Population: Similar to a genetic algorithm, microbial genetic algorithm has a population of potential solutions, but the population size tends to be smaller
    
    \item Fitness Function: Same as the conventional genetic algorithm.
    
    \item Tournament Selection: Two individuals are randomly selected from the selection for a tournament, with the less fit individual designated as the "loser" while the other is designated as the "winner".
    
    \item Crossover and Mutation: The genetic information of the winner is recombined and mutated before replacing the loser's genome. Only the losers of the tournaments are affected by the changes.
\end{enumerate}

\begin{table}[H]
\centering
\begin{tabular}{|c|c|}
\hline
\textbf{Individual} & \textbf{Trait} \\ \hline
\multirow{2}{*}{Doctor} 
 & Research Ability \\ \cline{2-2} 
 & Empathy \\ \hline
\multirow{4}{*}{Patient} 
 & Credibility Weight \\ \cline{2-2} 
 & Mean Rating Weight \\ \cline{2-2} 
 & Past Rating Weight \\ \cline{2-2}
 & Resilience \\ \hline
\end{tabular}
\caption{Traits used in the Microbial Genetic Algorithm for the classical simulation}
\label{tab:traits_classical}
\end{table}

\begin{table}[H]
\centering
\begin{tabular}{|c|c|}
\hline
\textbf{Individual} & \textbf{Trait} \\ \hline
\multirow{5}{*}{Doctor} 
 & Research Ability \\ \cline{2-2} 
 & Empathy \\ \cline{2-2} 
 & Social Ties \\ \cline{2-2}
 & Weight for Mean Weighted Rating \\ \cline{2-2}
 & Weight for Mean Weighted Respect \\ \hline
\multirow{5}{*}{Patient} 
 & Credibility Weight \\ \cline{2-2} 
 & Mean Rating Weight \\ \cline{2-2} 
 & Past Rating Weight \\ \cline{2-2}
 & Resilience \\ \cline{2-2}
 & Social Ties \\ \hline
\end{tabular}
\caption{Traits used in the Genetic Algorithm for the cognitive social systems simulation}
\label{tab:traits_cognitive}
\end{table}

Our implemented microbial genetic algorithm is a little different from usual because it uses elitism alongside tournament selection. Elitism is genetic algorithm strategy where one or more of the best individuals, also called elites, are guaranteed to carry over from the current generation to the next \cite{Katoch.etal2021}. Otherwise, the algorithm is similar to a convention microbial genetic algorithm. We used elitism and tournament selection, with five individuals being randomly selected instead of the usual two, after observing that without them, we were unable to increase the fitness for doctor and patient agents consistently in both of our models. The hyper-parameters for our implementations were chosen through trial and error, with fitness values for each agent being the key factor; we chose the hyper-parameters because they led to a higher, stable values for the fitness parameters. Our implementation of the genetic algorithm is representative of a co-evolutionary process \cite{Potter1997}, with both doctor and patient agents trying to reach their objectives given their interactions with each other. The relationship here between the agents is cooperative, meaning that one's attainment of objective does not hamper the other from achieving the same \cite{Popovici.etal2012}.  Traits or chromosomes for the doctor agent in the classical model are research ability and empathy. In addition to these traits, social ties for other agents and weights for calculating confidence are also included as traits for cognitive social system model. The traits for patient agents are resilience and weights for the judgement of doctors for the classical model, while social tie values are also included as traits for the cognitive social system model. Table \ref{tab:traits_classical} and Table \ref{tab:traits_cognitive} summarize the traits found for the classical and cognitive social system model respectively. Listing \ref{code:genetic_algo_overview} provides an outline of the methods for the microbial genetic algorithm implementation for the classical model. The cognitive social system model essentially has the same structure for the algorithm, except that the mutation chance is much lower at $0.01$, and the crossover chance is much higher at $0.5$. A more detailed overview for the components of the algorithm is given below. 

\begin{algorithm}[h]
\caption{Microbial Genetic Algorithm Operations in Classical Model}
\begin{algorithmic}[1]

\Function{SelectTwoIndividuals}{individuals}
    \State $tournament\_size \gets 5$
    \State $tournament \gets$ random.sample$(individuals, tournament\_size)$
    \State $tournament\_sorted \gets$ sorted$(tournament, key = \lambda x: x.calculate\_fitness(), \text{reverse} = \text{True})$
    \State \Return $tournament\_sorted[0], tournament\_sorted[-1]$
\EndFunction

\Function{Mutate}{individual}
    \State $mutation\_chance \gets 0.5$
    \If{random.random() $<$ mutation\_chance}
        \If{isinstance(individual, Patient)}
            \State individual.mutate\_judgment\_traits()
        \Else
            \State individual.mutate()
        \EndIf
    \EndIf
\EndFunction

\Function{Crossover}{individual1, individual2}
    \State $crossover\_chance \gets 0.3$
    \If{random.random() $<$ crossover\_chance}
        \State individual1.crossover$(individual2)$
    \EndIf
\EndFunction

\Function{PreserveElites}{individuals, num\_elites}
    \State $elites \gets$ sorted$(individuals, key = \lambda x: x.calculate\_fitness(), \text{reverse} = \text{True})[:num\_elites]$
    \State \Return $elites$
\EndFunction

\end{algorithmic}
\label{code:genetic_algo_overview}
\end{algorithm}

\subsubsection{Mutation}

The mutation component of the microbial genetic algorithm is different for the doctor and patient agent, and also it differs between the models. We will provide an overview of the differences below. 

In the classical model, for the doctor agent, the recent rating received from the patients dictate the change in mutation factor, which is the strength of the change mutation can induce in the doctor's traits. If the recent rating for the doctor is less than $3$, the mutation factor is $3$ and if not, it is $0.5$. We also make sure that the mutation factor does not exceed the personal resource and that if the value of the overall mutation change is $0.7$, we mutate the doctor's research ability. Otherwise, we mutate the doctor's empathy; so we mutate the research ability with a higher probability. We also ensure that the doctor's traits do not exceed their original range. In the cognitive social system model, the mutation method now include both weights that influence confidence score, weight for mean respect and weight for mean rating. It also includes a way to mutate social tie value of a random connection in the social network. Recent rating has less of an influence, with the mutation factor being $1.5$ if the the recent rating is $3$. Now, with equal $20\%$ chance, mutation occurs in all the traits. The bounds for the values are maintained in their original ranges and checked for consistency. Listings \ref{code:mutation_classical} and \ref{code:mutate_cognitive} demonstrate the implementation of mutation for the genetic algorithm in the simulation for the doctor agent. 

For the patient change, the mutation method for the microbial genetic algorithm are similar for both types of models. In the classical model, since the weights for the judgement score has to equal to $1$, the values for the weights are normalized so that their mutated values sum to $1$ and their bounds are checked so that their original range remains consistent. The weights and the resilience of the patient agent are mutated by a very small amount, with the mutation amount ranging from $(-0.05, 0.05)$. In addition to these traits, social strength value of one of the random connections of the patient is mutated with the chance of $50\%$. Listing \ref{code:mutate_patient} give the summary of mutation method for the patient agent in the cognitive social system model.

\begin{algorithm}[h]
\caption{Mutate Doctor Traits}
\begin{algorithmic}[1]

\Function{Mutate}{doctor}
    \State $mutation\_factor \gets$ recent feedback influences mutation strength
    \State $mutation\_amount \gets \min($doctor.personal\_resource$, random.uniform(0, 0.05)) \times mutation\_factor$
    \State $trait\_to\_mutate \gets random.random()$
    
    \If{$trait\_to\_mutate < 0.7$}
        \State $change \gets mutation\_amount \times random.choice([-1, 1])$
        \State doctor.research\_ability $\gets$ doctor.research\_ability $+$ change
        \State doctor.research\_ability $\gets \max(0, \min($doctor.research\_ability$, 1))$
    \Else
        \State $change \gets mutation\_amount \times random.choice([-1, 1])$
        \State doctor.empathy $\gets$ doctor.empathy $+$ change
        \State doctor.empathy $\gets \max(0, \min($doctor.empathy$, 1))$
    \EndIf
    \State doctor.personal\_resource $\gets$ doctor.personal\_resource $-$ mutation\_amount
\EndFunction

\end{algorithmic}
\label{code:mutation_classical}
\end{algorithm}

\begin{algorithm}[H]
\caption{Mutate Doctor Traits and Social Ties}
\begin{algorithmic}[1]

\Function{Mutate}{}
    \State $recent\_feedback \gets$ self.rating\_manager.get\_recent\_feedback(self.doctor\_id)
    \State $mutation\_factor \gets$ 1.5 \textbf{if} $recent\_feedback < 3$ \textbf{else} 0.5
    \State $mutation\_amount \gets$ random.uniform(0, 0.05) $\times$ mutation\_factor
    \State $trait\_to\_mutate \gets$ random.random()

    \If{$trait\_to\_mutate < 0.2$ \textbf{and} self.personal\_resource $> 0$}
        \State $change \gets$ mutation\_amount $\times$ random.choice([-1, 1])
        \If{$0 \leq$ self.research\_ability $+$ change $\leq$ 1}
            \State $actual\_change \gets$ min(abs(change), self.personal\_resource)
            \State self.research\_ability $\gets$ self.research\_ability $+$ actual\_change $\times$ (1 \textbf{if} change $>$ 0 \textbf{else} -1)
            \State self.personal\_resource $\gets$ self.personal\_resource $-$ actual\_change
        \EndIf
    \ElsIf{$trait\_to\_mutate < 0.4$ \textbf{and} self.personal\_resource $> 0$}
        \State $change \gets$ mutation\_amount $\times$ random.choice([-1, 1])
        \If{$0 \leq$ self.empathy $+$ change $\leq$ 1}
            \State $actual\_change \gets$ min(abs(change), self.personal\_resource)
            \State self.empathy $\gets$ self.empathy $+$ actual\_change $\times$ (1 \textbf{if} change $>$ 0 \textbf{else} -1)
            \State self.personal\_resource $\gets$ self.personal\_resource $-$ actual\_change
        \EndIf
    \ElsIf{$trait\_to\_mutate < 0.6$}
        \State $change \gets$ mutation\_amount $\times$ random.choice([-1, 1])
        \State self.weight\_wmrat $\gets$ max(0, min(1, self.weight\_wmrat $+$ change))
    \ElsIf{$trait\_to\_mutate < 0.8$}
        \State $change \gets$ mutation\_amount $\times$ random.choice([-1, 1])
        \State self.weight\_mwres $\gets$ max(0, min(1, self.weight\_mwres $+$ change))
    \Else
        \If{random.random() $<$ 0.5 \textbf{and} self.social\_ties\_doctors}
            \State $random\_doc\_id \gets$ random.choice(list(self.social\_ties\_doctors.keys()))
            \State $change \gets$ mutation\_amount $\times$ random.choice([-1, 1])
            \State self.social\_ties\_doctors[$random\_doc\_id$] $\gets$ max(0, min(1, self.social\_ties\_doctors[$random\_doc\_id$] $+$ change))
        \ElsIf{self.social\_ties\_patients}
            \State $random\_patient\_id \gets$ random.choice(list(self.social\_ties\_patients.keys()))
            \State $change \gets$ mutation\_amount $\times$ random.choice([-1, 1])
            \State self.social\_ties\_patients[$random\_patient\_id$] $\gets$ max(0, min(1, self.social\_ties\_patients[$random\_patient\_id$] $+$ change))
        \EndIf
    \EndIf

    \State self.personal\_resource $\gets$ max(0, self.personal\_resource) \Comment{Ensure resources do not drop below zero}
\EndFunction

\end{algorithmic}
\label{code:mutate_cognitive}
\end{algorithm}

\begin{algorithm}[H]
\caption{Mutate Patient Traits and Social Ties}
\begin{algorithmic}[1]

\Function{Mutate}{}
    \State $mutation\_amount \gets$ random.uniform(-0.05, 0.05)
    \State self.cred\_weight $\gets$ self.cred\_weight $+$ mutation\_amount
    \State self.mean\_rating\_weight $\gets$ self.mean\_rating\_weight $+$ mutation\_amount
    \State self.past\_rating\_weight $\gets$ self.past\_rating\_weight $-$ 2 $\times$ mutation\_amount

    \State $resilience\_change \gets$ random.uniform(-0.05, 0.05)
    \State self.resilience $\gets$ self.resilience $+$ resilience\_change
    \State self.resilience $\gets$ max(0.1, min(self.resilience, 0.4))

    \State $total \gets$ self.cred\_weight $+$ self.mean\_rating\_weight $+$ self.past\_rating\_weight
    \If{$total > 0$}
        \State self.cred\_weight $\gets$ self.cred\_weight / $total$
        \State self.mean\_rating\_weight $\gets$ self.mean\_rating\_weight / $total$
        \State self.past\_rating\_weight $\gets$ self.past\_rating\_weight / $total$
    \Else
        \State self.cred\_weight $\gets$ 1/3
        \State self.mean\_rating\_weight $\gets$ 1/3
        \State self.past\_rating\_weight $\gets$ 1/3
    \EndIf

    \State Ensure weights are within valid bounds
    \State self.cred\_weight $\gets$ max(0, min(self.cred\_weight, 1))
    \State self.mean\_rating\_weight $\gets$ max(0, min(self.mean\_rating\_weight, 1))
    \State self.past\_rating\_weight $\gets$ max(0, min(self.past\_rating\_weight, 1))

    \If{random.random() $<$ 0.5}  \Comment{Mutate social ties with doctors}
        \For{doc\_id in self.social\_ties\_doctors}
            \State $change \gets$ random.uniform(-0.1, 0.1)
            \State self.social\_ties\_doctors[doc\_id] $\gets$ self.social\_ties\_doctors[doc\_id] $+$ change
            \State self.social\_ties\_doctors[doc\_id] $\gets$ max(0, min(1, self.social\_ties\_doctors[doc\_id]))
        \EndFor
    \Else  \Comment{Mutate social ties with patients}
        \For{patient\_id in self.social\_ties\_patients}
            \State $change \gets$ random.uniform(-0.1, 0.1)
            \State self.social\_ties\_patients[patient\_id] $\gets$ self.social\_ties\_patients[patient\_id] $+$ change
            \State self.social\_ties\_patients[patient\_id] $\gets$ max(0, min(1, self.social\_ties\_patients[patient\_id]))
        \EndFor
    \EndIf
\EndFunction

\end{algorithmic}
\label{code:mutate_patient}
\end{algorithm}

\subsubsection{Crossover}

The crossover process for the doctor agent are similar across the models. In the classical model, we only focus on research ability and empathy and we take the average of these traits from the two selected individuals and create a new offspring with those values. In the cognitive social systems model, other traits are included and are similarly combined for the new offspring. Listing \ref{code:crossover_doctor} illustrates how the traits are utilized for the cognitive social system model. In the classical model, the crossover is performed with $30\%$ chance instead of $50\%$ as it is done for the cognitive social system model.

For the patient agent, the crossover process is once again similar for both types of models. With a $50\%$ chance, all the traits are averaged and passed of to the offspring from the parents in the previous generation. The bounds for the traits are once checked for consistency. Listing \ref{code:crossover_patient} depicts the process for the classical model; the cognitive social system model model has the same implementation along with the social tie values. 

\begin{algorithm} [h]
\caption{Perform Crossover Between Two Doctors}
\begin{algorithmic}[1]

\Function{Crossover}{other}
    \If{random.random() $<$ 0.5}  \Comment{Perform crossover with a 50\% chance}
        \State self.research\_ability $\gets$ (self.research\_ability + other.research\_ability) / 2
        \State self.empathy $\gets$ (self.empathy + other.empathy) / 2
        \State self.weight\_wmrat $\gets$ (self.weight\_wmrat + other.weight\_wmrat) / 2
        \State self.weight\_mwres $\gets$ (self.weight\_mwres + other.weight\_mwres) / 2

        \Comment{Averaging social ties between doctors, if applicable}
        \For{key in self.social\_ties\_doctors.keys()}
            \If{key in other.social\_ties\_doctors}
                \State self.social\_ties\_doctors[key] $\gets$ (self.social\_ties\_doctors[key] + other.social\_ties\_doctors[key]) / 2
            \EndIf
        \EndFor

        \Comment{Similar approach for patient social ties, if implemented}
        \For{key in self.social\_ties\_patients.keys()}
            \If{key in other.social\_ties\_patients}
                \State self.social\_ties\_patients[key] $\gets$ (self.social\_ties\_patients[key] + other.social\_ties\_patients[key]) / 2
            \EndIf
        \EndFor
    \EndIf
\EndFunction

\end{algorithmic}
\label{code:crossover_doctor}
\end{algorithm}

\begin{algorithm}[h]
\caption{Perform Crossover Between Two Patients}
\begin{algorithmic}[1]

\Function{Crossover}{other}
    \If{random.random() $<$ 0.5}  \Comment{Perform crossover with a 50\% chance}
        \State self.resilience $\gets$ (self.resilience + other.resilience) / 2
        \State self.cred\_weight $\gets$ (self.cred\_weight + other.cred\_weight) / 2
        \State self.mean\_rating\_weight $\gets$ (self.mean\_rating\_weight + other.mean\_rating\_weight) / 2
        \State self.past\_rating\_weight $\gets$ (self.past\_rating\_weight + other.past\_rating\_weight) / 2

        \State $total \gets$ self.cred\_weight + self.mean\_rating\_weight + self.past\_rating\_weight
        \If{$total > 0$}
            \State self.cred\_weight $\gets$ self.cred\_weight / $total$
            \State self.mean\_rating\_weight $\gets$ self.mean\_rating\_weight / $total$
            \State self.past\_rating\_weight $\gets$ self.past\_rating\_weight / $total$
        \Else
            \State self.cred\_weight $\gets$ 1/3
            \State self.mean\_rating\_weight $\gets$ 1/3
            \State self.past\_rating\_weight $\gets$ 1/3
        \EndIf


    \EndIf
\EndFunction

\end{algorithmic}
\label{code:crossover_patient}
\end{algorithm}

\subsubsection{Selection and Elitism}

The logic of selection and elitism is the same for all models and all the agents, as described in Listing \ref{code:genetic_algo_overview} that highlighted how the microbial genetic algorithm is implemented. 

\subsubsection{Fitness}

The fitness function remains unchanged for both agents across the models. For the doctor, the fitness function returns the mean rating of the doctor, as rated by all the patients who have been treated by the doctor for upto the amount of rounds that has passed. The Listing \ref{code:doctor_fitness} outlines the process. Similarly, for the patient agent, the fitness function returns the average health of the patient across all the rounds that have been experienced thus far. 

\begin{algorithm}[H]
\caption{Calculate Fitness of a Doctor}
\begin{algorithmic}[1]

\Function{CalculateFitness}{}
    \State $base\_fitness \gets 0$
    \If{self.rating\_manager}
        \State $base\_fitness \gets$ self.rating\_manager.get\_mean\_rating(self.doctor\_id)
    \EndIf
    \State \textbf{return} $base\_fitness$
\EndFunction

\end{algorithmic}
\label{code:doctor_fitness}
\end{algorithm}

\subsection{Simulation Logic}

The outline of the simulation logic of both models is given in Listing \ref{code:simulation_logic}. The main idea is that for each round, a certain number of patients are infected. These patients then seek out doctors, based on their judgment score and search criteria. The doctors then treat these patients with certain effectiveness, and subsequently, the genetic algorithm operations are performed for the individuals in both populations. 

The simulation is run for $100$ rounds, or mini-generations, for both models, and each round is repeated $50$ times. The number of doctors are $100$ and the number of patients are $1000$. The number of patients infected is $200$. We utilize two classes; rating manager and infection manager to manage rating calculations and spreading infections respectively for the simulation. However, we omit the details of these classes for the sake of brevity, as we have already included details about their utilities in our previous discussion about doctor and patient agents.

\begin{algorithm}[h]
\caption{Doctor-Patient Interaction in Simulation}
\begin{algorithmic}[1]
\State \textbf{Input:} $num\_doctors$, $num\_patients$, $num\_rounds$
\State Initialize $rating\_manager$
\State $doctors \gets$ initialize\_doctors($num\_doctors$, $rating\_manager$)
\State $patients \gets$ initialize\_patients($num\_patients$, $rating\_manager$)
\State $infection\_manager \gets$ new InfectionManager($patients$)
\For{$round \gets 1$ \textbf{to} $num\_rounds$}
    \State $infection\_manager$.spread\_infection()
    \State Reset $doctors$ to not busy
    \State $priority\_patients \gets$ sort $patients$ by priority
    \For{$patient$ in $priority\_patients$}
        \If{$patient$.needs\_doctor()}
            \State $chosen\_doctor \gets patient$.choose\_doctor($doctors$)
            \If{$chosen\_doctor$ \textbf{is not} $None$ \textbf{and not} $chosen\_doctor.is\_busy$}
                \State $patient$.receive\_treatment($chosen\_doctor$)
            \Else
                \State Log: Patient needs a doctor
            \EndIf
        \EndIf
    \EndFor
    \State Perform genetic operations on $patients$
    \State Perform genetic operations on $doctors$
    \State Calculate and log fitness metrics
\EndFor
\State \Return aggregated fitness and trait data over time
\end{algorithmic}
\label{code:simulation_logic}
\end{algorithm}

\vspace{3mm}

\clearpage

\section{Results and Analysis}

We ran the simulation as detailed in the previous section for our final result. However, we also simulated a single run to gain some insight and intuition on how the parameters or traits would behave. Also, we wanted to observe how the social network of agents change in the cognitive social systems model and it would be quite difficult to track the values over the multiple runs with our limited computational resources. Hence, we do not provide how the social ties change over the multiple runs of the simulation. Rather, we provide two graphical transformations of a patient agent and a doctor agent to demonstrate that in cognitive social system, reaching objectives for both agents does not necessarily mean tweaking the traits as encoded in the classical model. First, we will provide the results for a single run of both types of models. Then we will provide the evolution of traits for both agents over the multiple runs and compare the results. Finally, we will provide social network graphs and analyze how social agents utilise social ties to reach their goals. 

\subsection{Single Run}

For a single run of the simulation, we reduced the number of agents in the model. The number of doctors was reduced to $15$, the number of patients reduced to $100$ and rounds to $20$. The number of infected patients in the simulaltion was equal to the number of patients in the system. 

\begin{figure}[h!]
    \centering
    \includegraphics[width=1\linewidth]{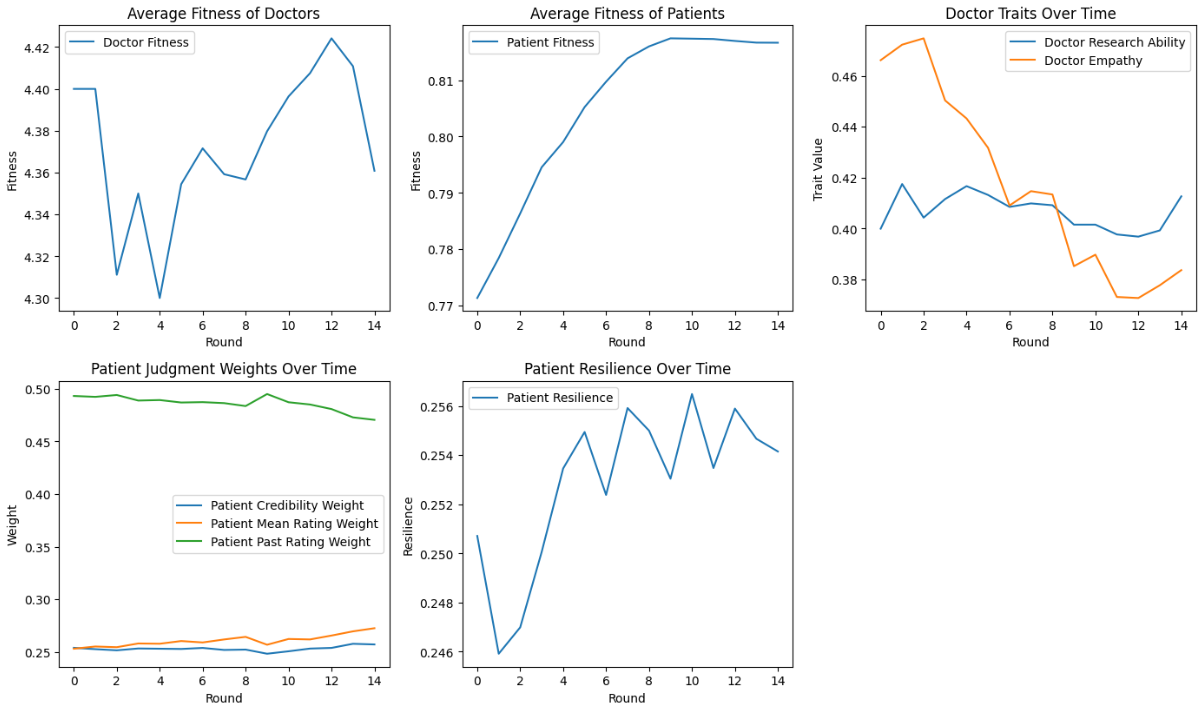}
    \caption{Single Run for Classical Model}
    \label{fig:classical_single_run}
\end{figure}

\begin{figure}[H]
    \centering
    \includegraphics[width=1\linewidth]{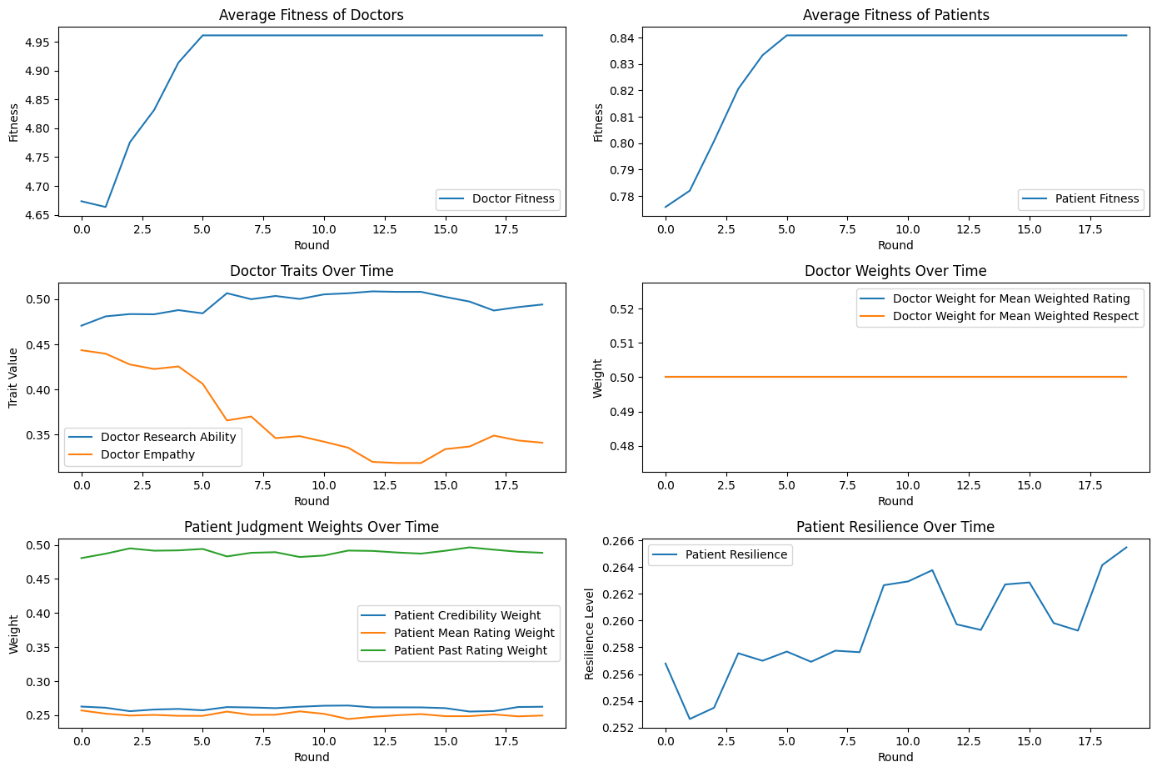}
    \caption{Single Run for Cognitive Social System Model}
    \label{fig:cog_social_single_run}
\end{figure}

We present the results for classical and cognitive social system model in Figure \ref{fig:classical_single_run} and Figure \ref{fig:cog_social_single_run} respectively. Looking at the figures, it appears that the doctor agent struggles to achieve fitness over the short number of rounds in the classical model while in the cognitive social system (CSS) model, a stable fitness value is reached quite quickly. The value of the fitness for the doctor agent also seem higher in the CSS model. The patient agent's fitness appear to be of the same amount for both models, and the fitness is discovered almost around the same time in both models. For the doctor agent, research ability and empathy show similar trends in both models; research ability seem to increase while empathy continues to evolve towards a lower value. The weights for the judgement of doctor for the patient seem relatively unchanged, and so do the weights for doctor's confidence. However, it is surprising to see that the resilience of the patient agent in both model shows an increasing trend, given that lowering resilience actually increases treatment effectiveness in our simulation. 

\subsection{Multiple Runs}

For the multiple runs, we first provide an overview of the final values for the simulation of both models, given in Table \ref{tab:simulation_metrics_classical} and Table \ref{tab:simulation_metrics_cognitive}. Then we will look at the fitness values for both doctors and patients, for both types of models and then we will look the evolution of the traits for both social agents.

\subsubsection{Final Values}

Looking at the final mean values for both types of models, it looks like both doctor and patient agents are able to reach a higher fitness value in the CSS model. However, the research ability of the doctor agent is considerably lower in the CSS model compared to the classical model, hinting that social tie evolution might be involved in improving the fitness values. The patient values his own past rating in the CSS model more, and patient's resilience in both models seem to be increasing. 

\begin{table}[H]
\centering
\begin{tabular}{@{}lcc@{}}
\toprule
\textbf{Metric} & \textbf{Mean (Final Round)}\\ \midrule
Doctor Fitness & 4.469  \\
Patient Fitness & 0.812  \\
Doctor Research Ability & 0.420\\
Doctor Empathy & 0.358 \\
Patient Credibility Weight & 0.254 \\
Patient Mean Rating Weight & 0.253 \\
Patient Past Rating Weight & 0.493  \\
Patient Resilience & 0.252\\ \bottomrule
\end{tabular}
\caption{Final Classical Model Simulation Metrics for Doctors and Patients }
\label{tab:simulation_metrics_classical}
\end{table}

\begin{table}[H]
\centering
\begin{tabular}{@{}lcc@{}}
\toprule
\textbf{Metric} & \textbf{Mean (Final Round)}\\ \midrule
Doctor Fitness & 4.972\\
Patient Fitness & 0.836\\
Doctor Research Ability & 0.405\\
Doctor Empathy & 0.360\\
Doctor Weight for Mean Weighted Rating & 0.500\\
Doctor Weight for Mean Weighted Respect & 0.500\\
Patient Credibility Weight & 0.251 \\
Patient Mean Rating Weight & 0.249 \\
Patient Past Rating Weight & 0.499 \\
Patient Resilience & 0.262\\ \bottomrule
\end{tabular}
\caption{Final Cognitive Social System Model Simulation Metrics for Doctors and Patients}
\label{tab:simulation_metrics_cognitive}
\end{table}

\subsubsection{Doctor Fitness}

\begin{figure}[H]
    \centering
    \includegraphics[width=1\linewidth]{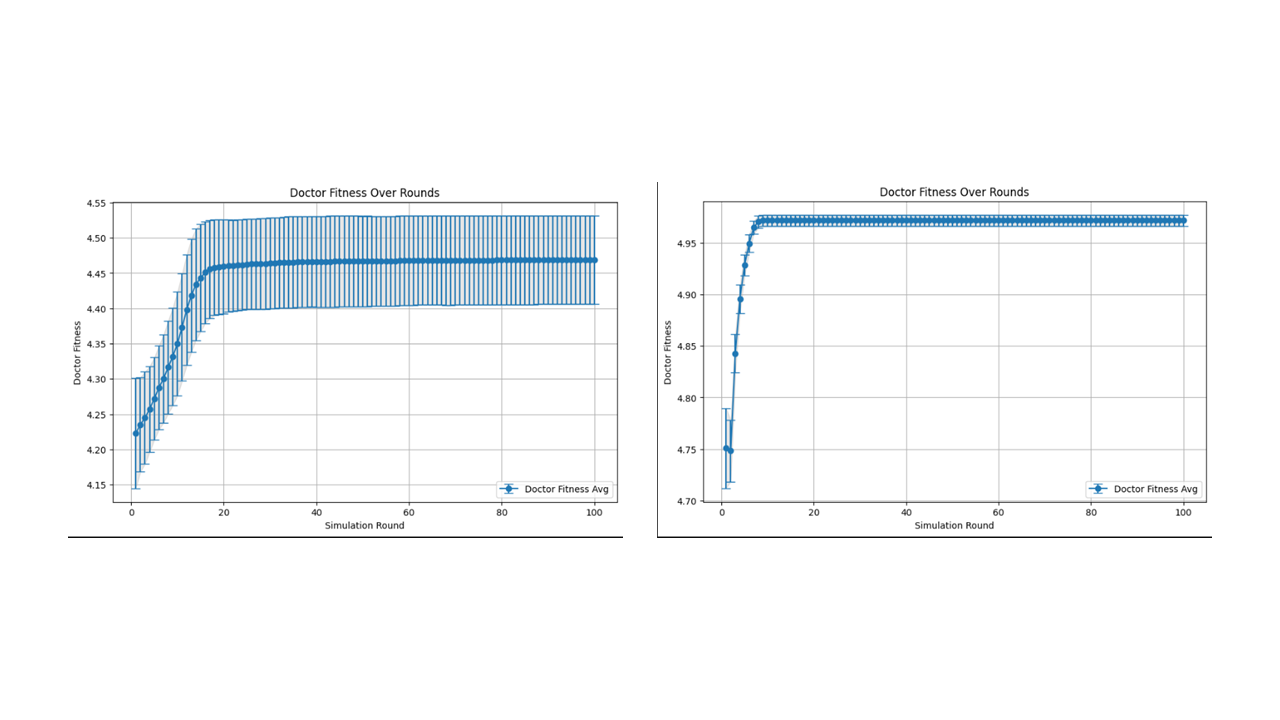}
    \caption{Fitness Evolution of Doctor Agent}
    \label{fig:doctor_fitness}
\end{figure}

For the fitness value of the doctor agent, provided in Figure \ref{fig:doctor_fitness}, it seems that the agent reaches a higher value faster in the CSS model. This is despite the added complexity the CSS model integrates in its structure. However, the added complexity also means more feedback for the agent, and thus could be construed as a more enabling model. 

\subsubsection{Traits for the Doctor Agent}

\begin{figure}[H]
    \centering
    \includegraphics[width=1\linewidth]{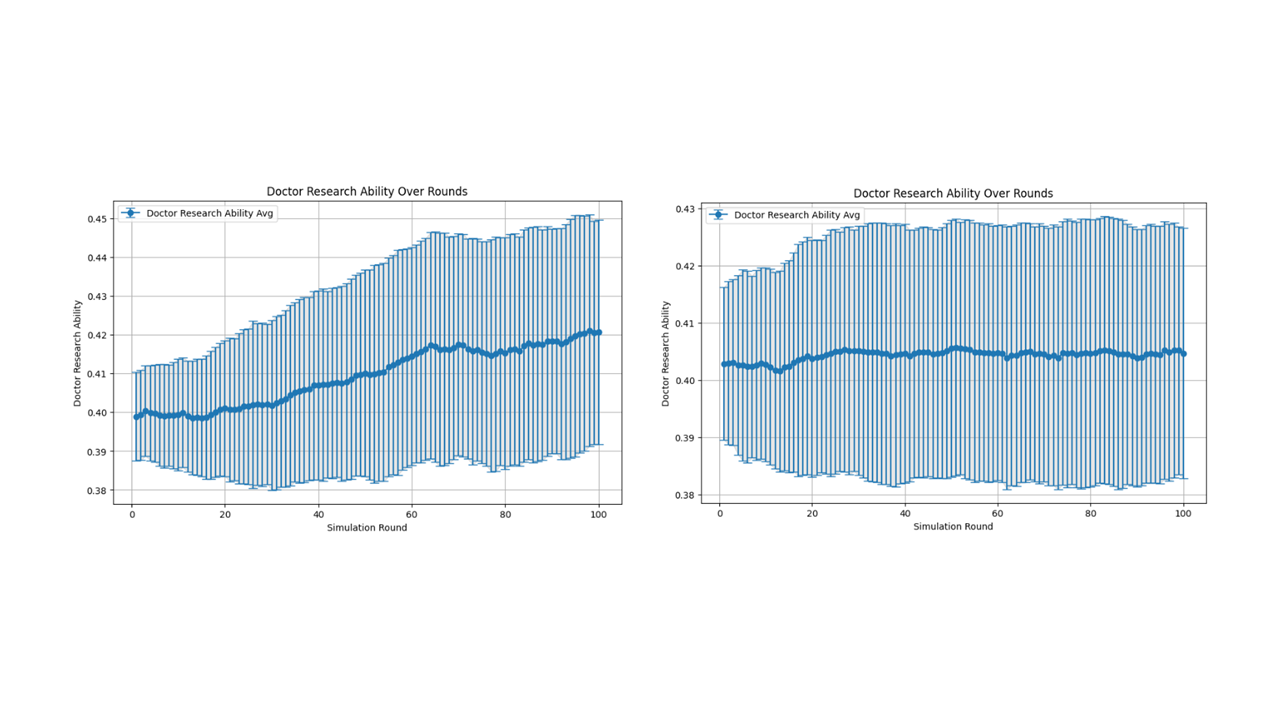}
    \caption{Evolution of Doctor's Research Ability}
    \label{fig:research_ability}
\end{figure}

\begin{figure}[H]
    \centering
    \includegraphics[width=1\linewidth]{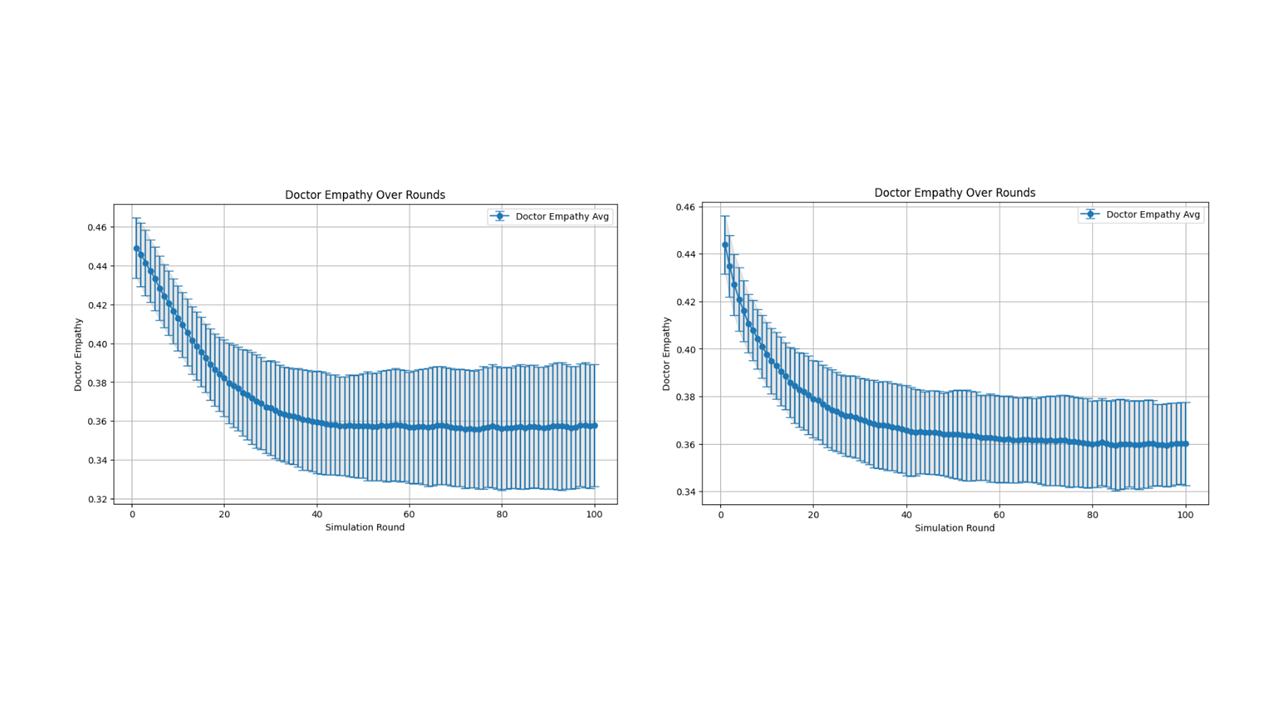}
    \caption{Evolution of Doctor's Empathy}
    \label{fig:empathy}
\end{figure}

\begin{figure}[H]
    \centering
    \includegraphics[width=1\linewidth]{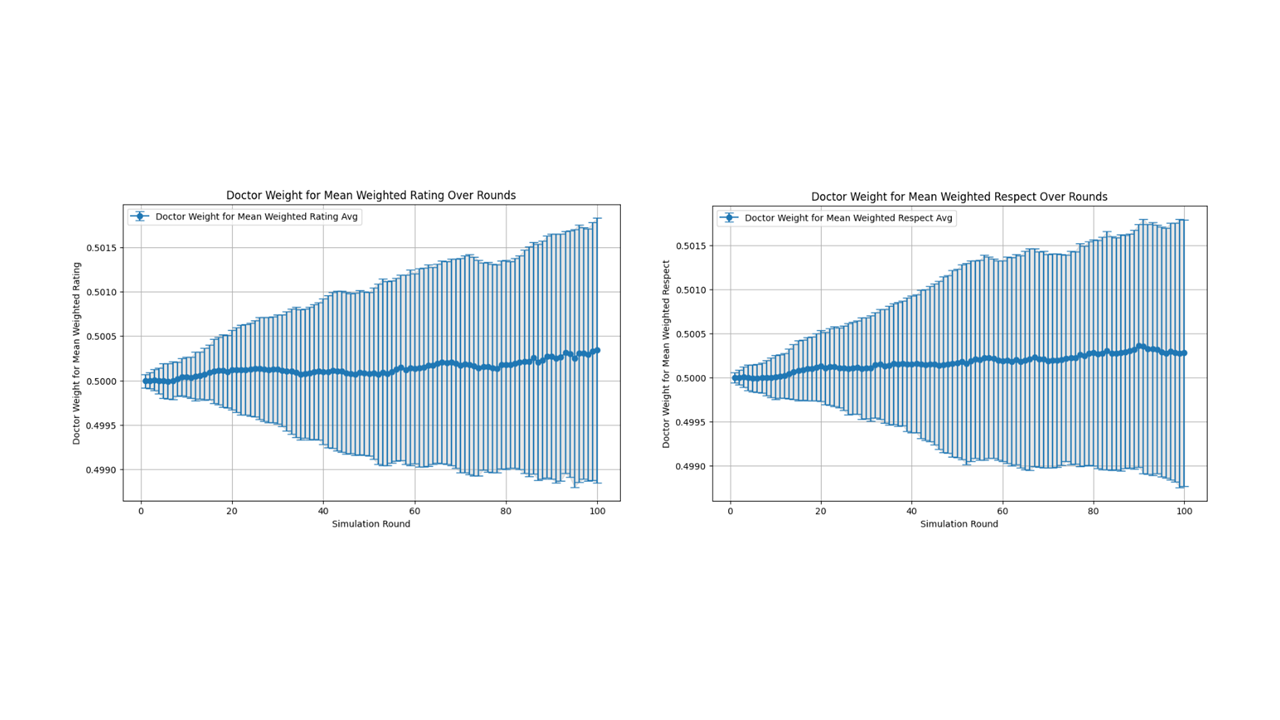}
    \caption{Evolution of Doctor's Confidence Weights}
    \label{fig:confidence weights}
\end{figure}

The graphs for the doctor's research ability, empathy and weights for confidence are presented in Figure \ref{fig:research_ability}, Figure \ref{fig:empathy} and Figure \ref{fig:confidence weights}, respectively. The weights for the confidence components do not seem to change much at all, either because the weights are already initialized at an optimum level or because other traits, such as research ability, drive the increase in fitness. However, looking at the research ability of the doctor agent, we note that the research ability does not show a sufficient increase in the CSS model. In contrast, it shows steady growth in the classical model. The consistent decrease in empathy is also fascinating since it is one of the components that determine treatment effectiveness, the average health level of patients, and the value of the rating of the doctor agent. In the classical model, this decreasing trend in empathy can be explained by increased research ability since personal resources constrain both. We are not sure why this occurred in the CSS model, which means either the agent compensates for this decrease through social ties or our model has a structural problem.

\subsubsection{Patient Fitness}

\begin{figure}[H]
    \centering
    \includegraphics[width=1\linewidth]{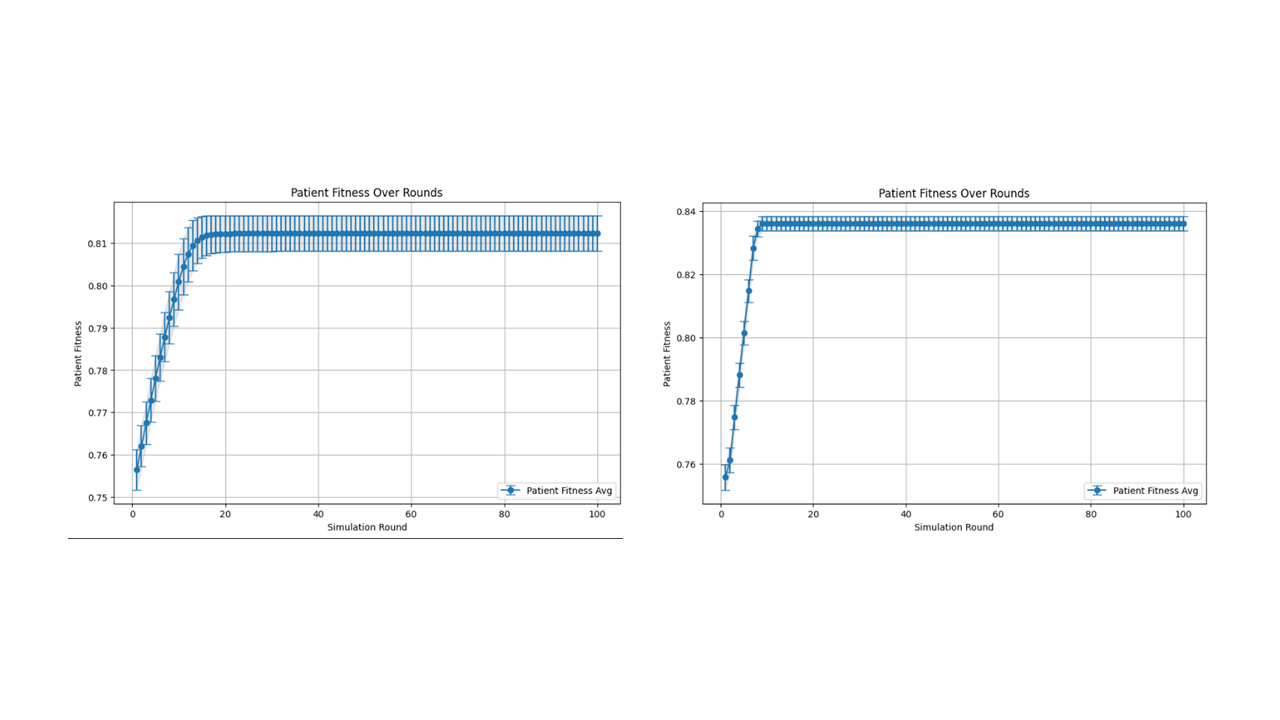}
    \caption{Fitness Evolution of Patient Agent}
    \label{fig:patient_fitness}
\end{figure}

Figure \ref{fig:patient_fitness} provides a look at the evolution of patient's fitness values over the rounds. In the CSS model, similar to the doctor agent, the patient agent is able to achieve a higher fitness value in a relatively shorter period of time than in the classical model. 

\subsubsection{Traits for the Patient Agent}

\begin{figure}[h]
    \centering
    \includegraphics[width=1\linewidth]{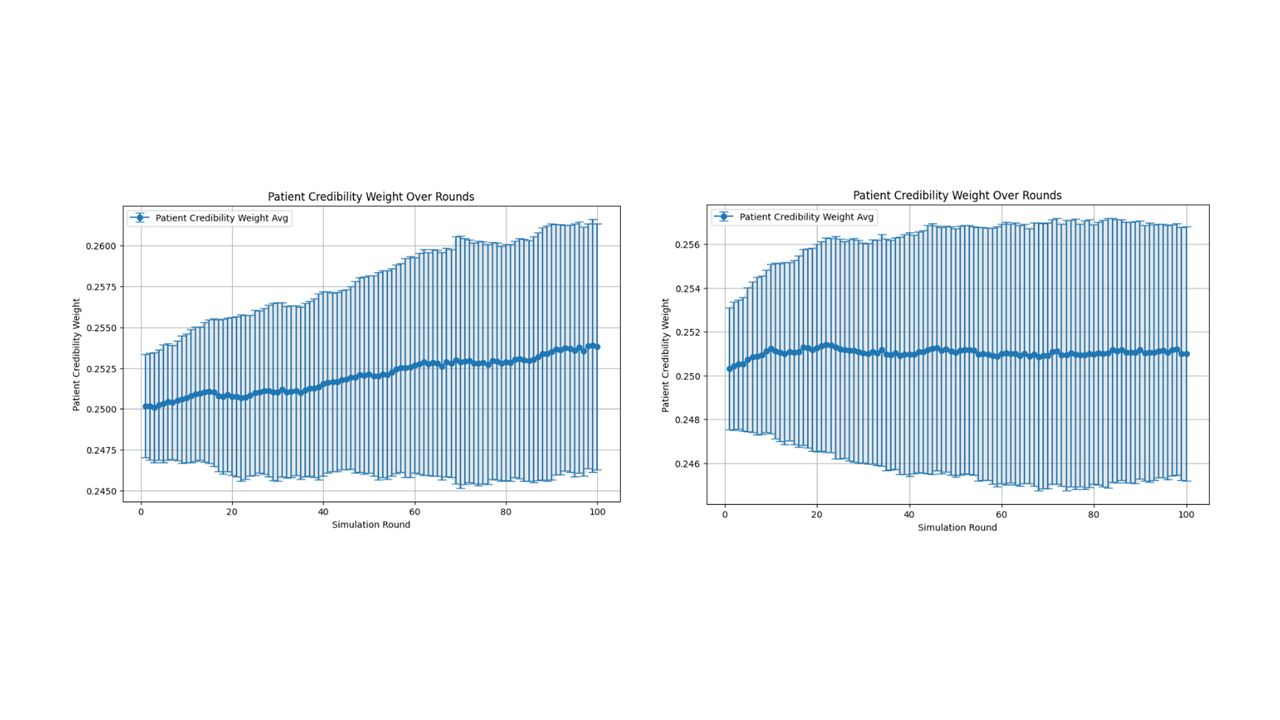}
    \caption{Evolution of Patient's Weight for Credential}
    \label{fig:patient_credential}
\end{figure}

\begin{figure}[h]
    \centering
    \includegraphics[width=1\linewidth]{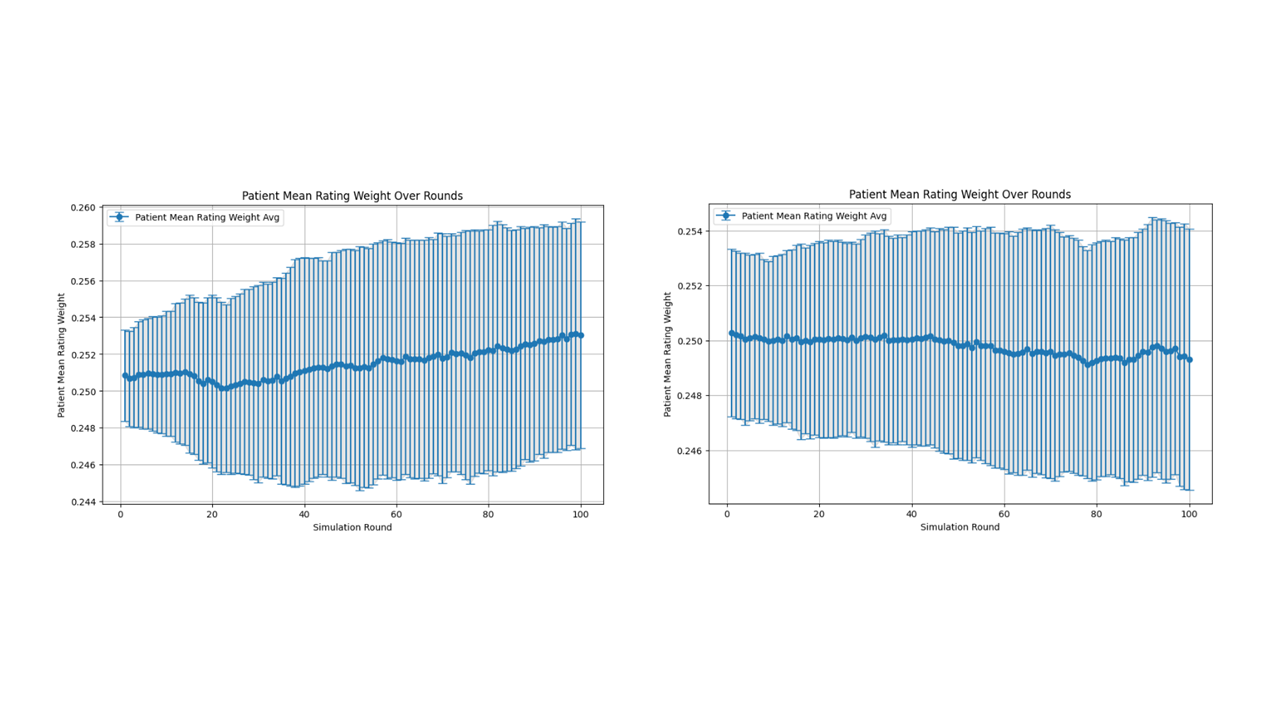}
    \caption{Evolution of Patient's Weight for Mean Rating}
    \label{fig:patient_mean_weight}
\end{figure}

\begin{figure}[h]
    \centering
    \includegraphics[width=1\linewidth]{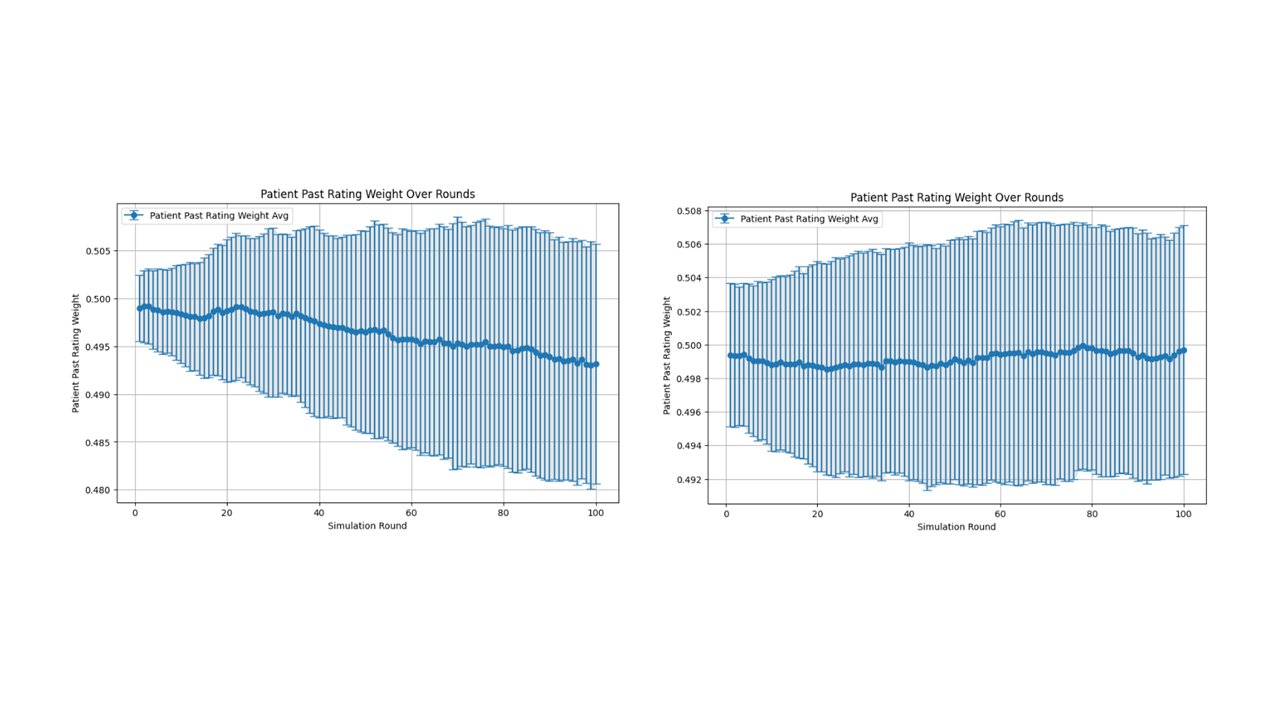}
    \caption{Evolution of Patient's Weight for Past Rating}
    \label{fig:patient_past_rating}
\end{figure}

\begin{figure}[h]
    \centering
    \includegraphics[width=1\linewidth]{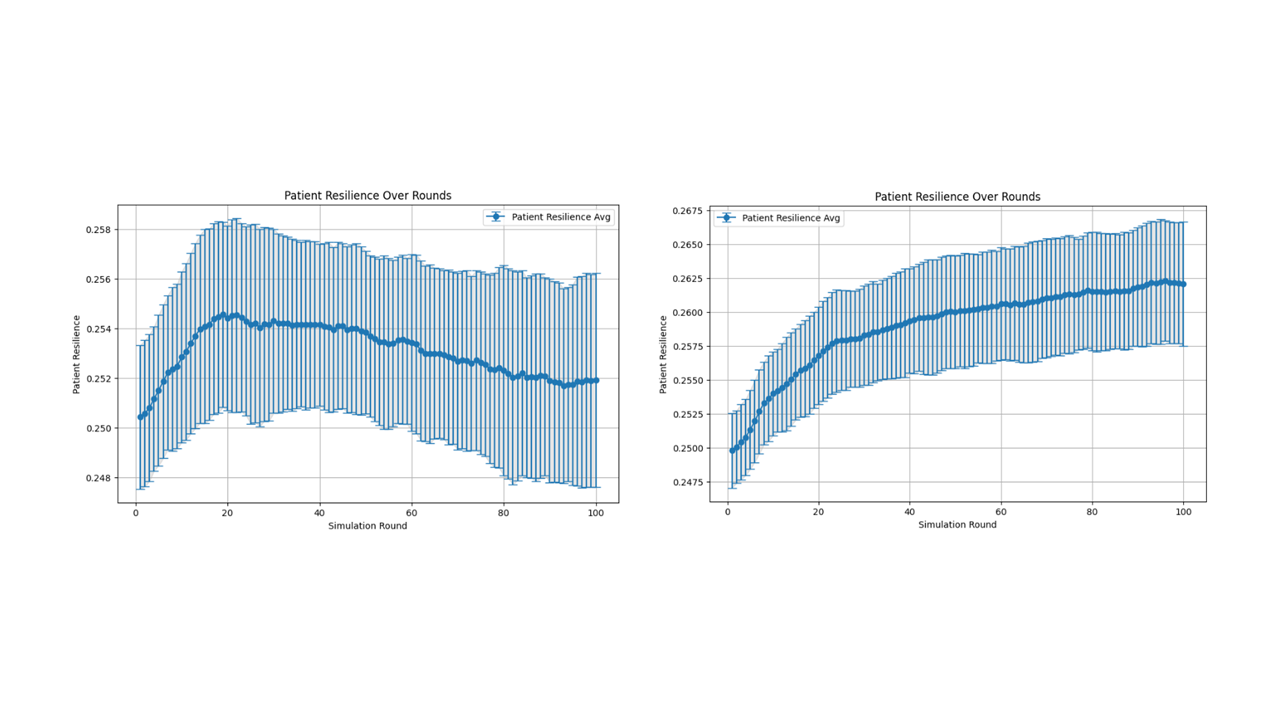}
    \caption{Evolution of Patient's Resilience}
    \label{fig:patient_resilience}
\end{figure}

The traits of the patient agent evolve relatively differently for the two models. First, the patient's resilience steadily increases in the CSS model (See Figure \ref{fig:patient_resilience}), whereas in the classical model, the trait gradually decreases. We expected this trait to show a decreasing trend, given that it lowers treatment effectiveness. Thus, it was surprising to find an increasing trend in the CSS model. The weights for the patient's judgement of the doctor are comparatively stable. At the same time, it shows a slight increase for the patient's past rating while showing a slight decrease in mean rating for the CSS model in Figure \ref{fig:patient_past_rating} and Figure \ref{fig:patient_mean_weight} respectively, the increase in values are pretty minute. However, for the classical model, the weights for credentials, as displayed in Figure \ref{fig:patient_credential}, and mean rating show an increasing trajectory. In contrast, the weight on past ratings shows a steady decline.

\subsection{Social Network Graphs for Cognitive Social System Model}

\begin{figure}[h]
    \centering
    \includegraphics[width=1\linewidth]{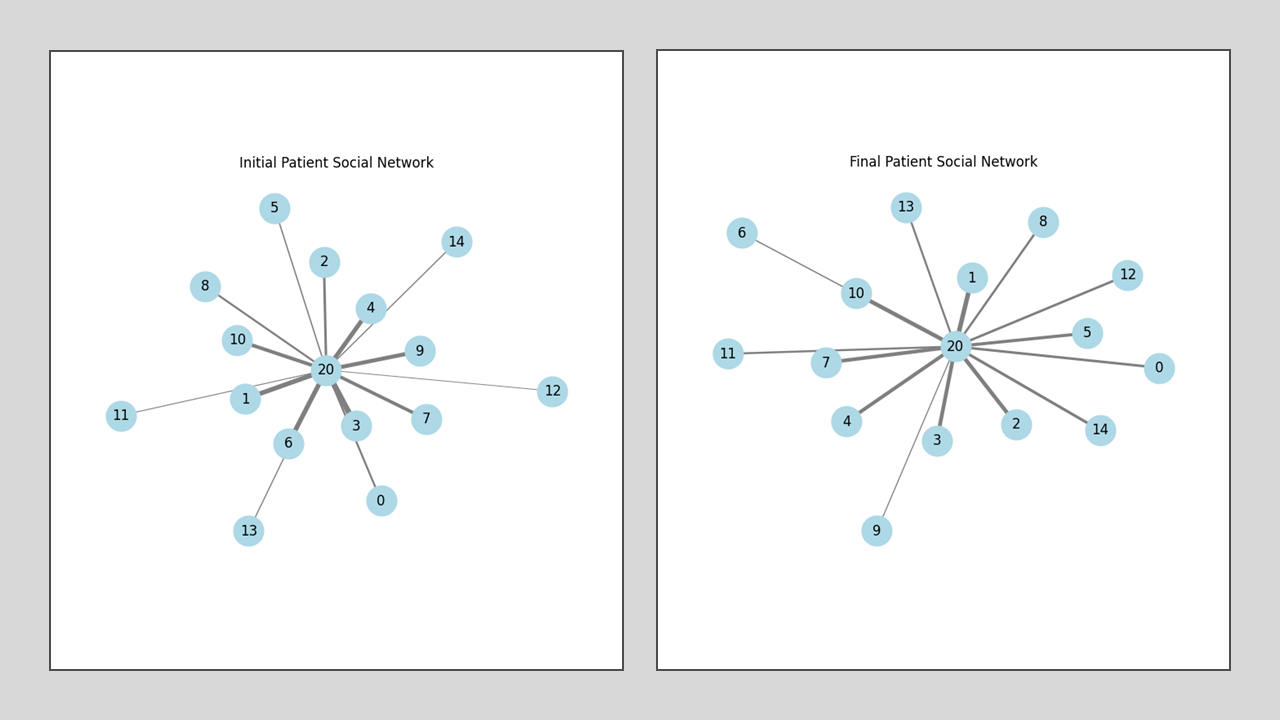}
    \caption{Patient Social Network Evolution}
    \label{fig:patient_social_network}
\end{figure}

\begin{figure}[h]
    \centering
    \includegraphics[width=1\linewidth]{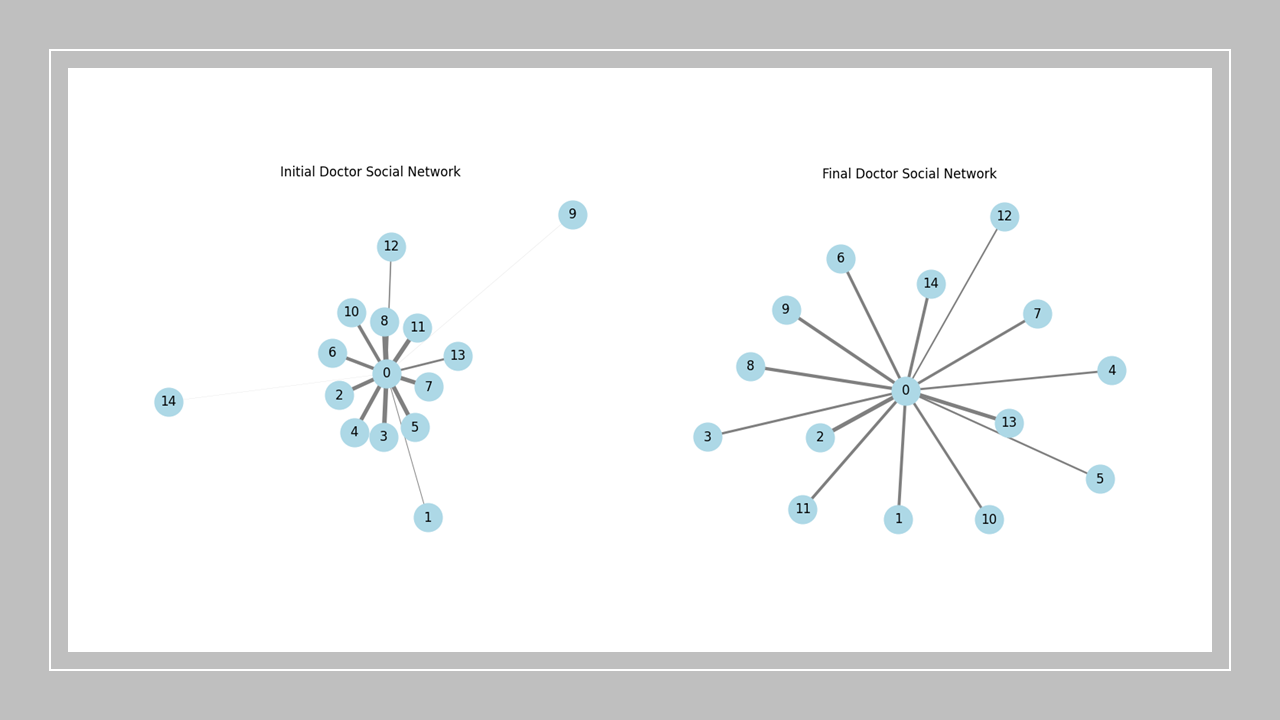}
    \caption{Doctor Social Network Evolution}
    \label{fig:doctor_social_network}
\end{figure}

The social network graphs given in Figure \ref{fig:patient_social_network} and Figure \ref{fig:doctor_social_network} are derived from a single run of the simulation where the number of doctors, patients and rounds are $15$, $100$ and $20$ respectively, similar to the conditions for results of the single run of the simulation presented earlier in this section. Despite their limited number of rounds, we observe that the transformations of the social network for the patients and doctors shed light on how each agent might be optimising their objectives. Both agents create more decisive social ties as their traits evolve through the rounds. For the doctor agent, it seems that she is strengthening her social ties with other doctors, thereby increasing confidence and, thus, the treatment effectiveness, leading to better ratings. As for the patient agent, creating stronger social ties might mean more reliable information about the doctor's treatment effectiveness and, thus, better average health level. The trends in these graphs also explain how the doctor agent can lower both research and empathy values and yet achieve better ratings; with stronger, favourable social ties, the doctor can effectively have higher, inflated ratings, given that ratings are modulated by the strength of the social ties, despite not improve her abilities.

\clearpage

\section{Discussion}

In this section, we will first discuss the consequences of our simulation results. Then, we will look at some limitations that restrict the generalisation and interpretation of the results we have derived in the previous section. Finally, we will discuss some of the steps that can be taken in the future to extend and develop more confidence in our modelling approach. 

While the results observed in the previous section were consistent with our expectations, we were still surprised by some. In particular, we had concluded that the doctor's research ability will remain the same in the CSS model. However, after looking at the changes taking place in the social networks of both doctors and patients, we realised that higher values in social ties might be enough to lead to an increase in confidence through an increase in ratings from patients and respect values from other doctors, and ratings from the patient, through an increase in treatment effectiveness. Thus, an increase in social tie strengths forms a cycle of positive ratings being achieved by the doctor. What surprised us the most was the steady decrease in empathy in both models for the doctor agent and the increase in resilience for the patient agent in the CSS model. This decreased empathy could indicate a shift in the doctor's focus from emotional connection to professional interest. A decline in empathy values might be due to the doctor agent prioritising research ability to increase credentials and, therefore, improve treatment effectiveness through that path as evident in the classical model (See Figure \ref{fig:research_ability}). On the other hand, the increase in resilience for the patient agent could suggest a higher tolerance for treatment, leading to better health outcomes. A lower value for resilience is better for patients since it increases the effective treatment received from doctors. However, higher resilience means that patients find doctors who increase their health levels better than others in the simulation. Thus, it can be an indirect way of finding out which doctors are the most effective. 

Our approach to modelling the simulation adds some inherent stochastic elements to both models, thus creating a much more dynamic process for the interactions between the social agents. The infection affects patients randomly, while research abilities, empathy, resource constraints and resilience are also endowed randomly. In the CSS model, social ties are arbitrarily distributed at first. It was, therefore, interesting to observe that the agents attempted to make things more specific and deterministic by increasing their social tie strengths in the CSS model. By developing stronger bonds, the doctor agent tried to improve her rating. In contrast, the patient agent did the same to increase the health level since their objectives are interdependent in both models.  

Our decision to design the models from the ground-up, using information from relevant literature, enabled us to approach the modelling without any established bias about the nature of the healthcare system which can serve as foundational models for further research \cite{Tracy.etal2018}. Our approach can help policymakers find new leverage points \cite{Fischer.Riechers2019}, pinpointing specific areas to change in order to change the social dynamics and effectively intervene for desired results. Thus, the analysis presented in this paper is an attempt to understand the cognitive effects on social phenomena. Cognitive social system modelling can likely help us explain a range of social behaviours; the doctor agent's lack of research ability in the model can be explained by the density of the social network and the steady availability of patients despite low credentials. It can help us explain how qualified professionals are often motivated by political and social influences and exhibit multiple mechanisms that can lead to the same goals or outcomes \cite{Davahli.etal2020}. The model can be modified for more complex, dynamic situations in the healthcare setting. The general principles based upon which the model was created can be applied to social systems other than healthcare.

\subsection{Limitations}

\subsubsection{Simulation}

Our approach to modelling is not without its limitations. In this section, we highlight some of them so that further work based on our research can be executed with more efficiency and applied to other areas with precise and accurate modifications. 

We note that the social agents created in our model display a level of homogeneity that is most likely to be absent from the physical world. 
For example, to begin the search for a doctor, a patient agent waits until the health level falls below a particular value, $0.6$. However, not all patients will have a threshold value of $0.6$ in the real world; a patient might be more risk averse and start searching for a doctor even when he is relatively healthy. We had also assumed that patients can make choices regarding their doctors, and that they are able to access the services of doctors without any restrictions. This aspect might not be accurate since the doctors themselves might have certain preferences or follow certain guidelines that might prevent the patients from choosing them \cite{Farago.etal2022}. Doctors may even choose to see a patient because of recommendation or social connection; this aspect is not captured by our models. 

Furthermore, we had also assumed that there are no correlations between the credential a doctor has and resource constraints associated with the doctor. However, it is likely that resource constraints become lower as the doctor advances in his career and gain higher credentials. Additionally, in our model, personal resource determines how much research ability and empathy can be improved. It can be thought of as a cognitive resource that can be utilized for either of the two attributes to modulate treatment effectiveness. It is fixed for every doctor to be 0.2. However, this is unrealistic because a doctor's personal resource might vary based on factors such as credential and geographic location among others \cite{Eklund2008}. Specialties among the doctors can also vary, leading to ineffective treatments due to ineffective matching with the patients.

Infection dynamics in the models could have been more complex to reflect the observations in the real world; recovery time of the patients may vary depending on the severity of the infections and the doctors themselves can get infected and thus unable to treat other patients. Patients themselves can have resource constraints; geographical limitations can prevent a patient from seeing the desired doctor. There could be time constraints that can prevent optimal search, such as an accelerate rate of deterioration due to infection. In our models, we have implemented a variation of loyalty among the patients; if the patient receives a satisfactory treatment, the patient will go back to the same doctor. This loyalty dynamic is rather simplistic and we usually see a variety of loyalty mechanisms that determine patient loyalty and desirability in the physical world \cite{Harris2003}.

We would also like to mention that we included variables like confidence, empathy and patient judgement, which are difficult to measure and assess due to their psychological nature. While measures such psychometric tests exist for these psychological traits and abilities \cite{Hojat.etal2002}, they are not easily amenable to direct interpretations or numerical encoding. Therefore, it would be prudent to use these variables once one is certain about their accuracy; verifying the values for these variables could involve statistical tests and conducting thorough observations or experiments.

\subsubsection{Learning Algorithm}

We would also like to point out that although we utilized a microbial genetic algorithm for the social agents in our models, we could not conduct a systematic parameter sweep to discover if there were other hyper parameters that could have led to better results. Our trial and error method is not efficient and it would be judicious to find a better way to identify the ideal parameters for the genetic algorithm. Additionally, since our algorithm converged quite fast to stable results, it is likely that the populations in the simulation were not exposed to enough variety through the genetic algorithm. Although the results had come out poorer without elitism and with higher chances of mutation, trying to find an alternative route might enable more variation and genetic diversity in the simulated individuals \cite{Katoch.etal2021}. Different algorithms such as reinforcement learning or other predictive models can also be utilized for the simulation in an attempt to find the most optimum outcomes for the social agents.

\subsection{Future Directions}

One way to improve upon our research would involve adapting the model to more complex scenarios such as emergency healthcare network, where there would be more than just doctor and patient agents \cite{Cabrera.etal2011}. It would involve setting up multiple objectives for each agent, with more complex relationships between doctors and patients and other stake holders in the healthcare industry \cite{Davahli.etal2020}. It would also be a good way to address the limitations presented earlier in this section; by incorporating complexities in other settings, the model becomes more robust.   

To improve upon the existing scenario of modelling primary healthcare network, modelling and collection of detailed metrics such as the average time a patient remains infected, the spread of infection rates, and other health statistics over multiple simulation runs can be a great way to enhance the analysis of results. Furthermore, adding dynamic doctor skills, where doctors are modelled to improve their skills or experience levels based on the number and type of cases they handle, which could affect their future ratings and choices by patients, can help add real world complexities to the model. Implementing more sophisticated algorithms for patient-doctor matching, potentially incorporating machine learning techniques to predict the best match based on past data, could enable others to use the models as a simulation for real world scenarios. Finally, it is important to point out that collecting data and seeing if the predicted results from the model line up with the values can be good starting point to extend the model and find key ingredients that can help generalize the model to different settings.

\clearpage

\printbibliography[heading=bibintoc, title={References}]

\end{document}